\documentclass[11pt,]{article}
 \usepackage{graphicx}
 \usepackage[cp1251]{inputenc}  
\begin{document}

 \bigskip

 \bigskip

 \centerline {\Large\bf Analysis of Orbital Dynamics of Globular Clusters}
 \centerline {\Large\bf in the Central Region of the Milky Way}

  \bigskip

  \bigskip

   \centerline{\bf
            A.~T.~Bajkova\footnote [1]{bajkova@gaoran.ru} (1), A.~A.~Smirnov (1),
            V.~V.~Bobylev (1)
            }
  \bigskip
  \bigskip
 \centerline{\small \it (1)Central (Pulkovo) Astronomical Observatory RAS, St. Petersburg, 196140 Russia}
  \bigskip

  \bigskip

{{\bf Abstract.}

The regularity/chaoticity of orbits of 45 globular clusters in the central region of the Galaxy with
a radius of 3.5 kpc, which are subject to the greatest influence of the elongated rotating bar, is analyzed.
Various methods of analysis are used, namely, the methods of calculating the maximum characteristic Lyapunov
exponents (MCLE), MEGNO (Mean Exponential Growth factor of Nearby Orbits), the Poincar\'e
section method, the frequency method based on calculating fundamental frequencies, and a new method
is proposed based on calculating the orbit amplitude spectrum as a function of time and calculating the
entropy of the amplitude spectrum as a measure of orbital chaos. Bimodality is found in the histogram of
the distribution of positive Lyapunov exponents calculated in the classical version, without renormalizing
the shadow orbit, which allows implementing a probabilistic method for GC classification, which is also a
new approach. To construct the orbits of globular clusters, we used the gravitational potential model with
a bar in the form of a triaxial ellipsoid. The following bar parameters were adopted: mass $10^{10} M_\odot$, length
of the semi-major axis 5 kpc, angle of rotation of the bar axis 25$^o$, rotation velocity 40 km s$^{-1}$ kpc$^{-1}$. To
form the 6D-phase space required for integrating the orbits, we used the most accurate astrometric data to
date from the Gaia satellite (EDR3), as well as new refined average distances to globular clusters. Globular
clusters with regular and chaotic dynamics were classified. As the analysis showed, globular clusters with
small pericentric distances and large eccentricities are most susceptible to the influence of the bar and
demonstrate the greatest chaos. It is shown that the results of the classification of globular clusters by
the nature of their orbital dynamics, obtained using the various analysis methods considered in the work,
correlate well with each other.
}

\bigskip


\section{INTRODUCTION}
This work continues a series of works by the authors
(Bajkova and Bobylev, 2022; Bajkova et al.,
2023a, b) devoted to the study of the orbital dynamics
of globular clusters (GCs). Thus, in the work of
Bajkova and Bobylev (2022) a catalog of orbits of
152 galactic GCs is presented based on the latest
astrometric data from the Gaia (Gaia EDR3) satellite
(Vasiliev and Baumgardt, 2021), as well as new
refined average distances (Baumgardt and Vasiliev,
2021). Based on the same data, in the work of
Bajkova et al. (2023a) an analysis of the influence of
the galactic bar on the orbital motion of GCs in the
central region of the Galaxy was carried out. For this
task we selected 45 GCs in the central galactic region
with a radius of 3.5 kpc, 34 of which belong to the
bulge and 11 to the disk. We obtained the GC orbits
both in an axisymmetric potential and in a potential
including a bar model in the form of a triaxial ellipsoid.
In this case, the mass, rotation velocity and dimensions
of the bar were varied. A comparison of such
orbital parameters as the apocentric and pericentric
distances, eccentricity and maximum distance from
the galactic plane was carried out.

The second stage of research aimed at studying
the influence of the bar on the orbital motion of the
GCs was devoted to the problem of identifying objects
captured by the bar using spectral dynamics methods
(Bajkova et al., 2023b).

The aim of this work is to analyze the regularity/chaoticity of the orbits of all 45 GCs, previously
selected in the central region of the Galaxy, using
various methods.

It should be noted that the problem of regularity/chaoticity of the orbital motion of GCs in the
central region of the Galaxy has already been considered
by us in the work of Bajkova et al. (2023b),
but this was done very superficially and based only on
the method of calculating the MCLE in the classical
version, i.e. without renormalizing the shadow orbit
obtained by perturbing the phase initial point. The
conclusions made by Bajkova et al. (2023b) are not
correct for all GCs in the sample, and a more in-depth
analysis is needed using several of the most effective
methods for estimating the chaotic nature of orbits.

Since the GCs in the central region of the Galaxy
are subject to the greatest influence from the elongated
rotating bar, the question of the nature of the
GC orbital motion -- regular or chaotic -- is of great
interest. For example, Machado and Manos (2016)
showed that the main share of chaotic orbits should
be in the bar region.

In this paper, we limit ourselves to considering
the problem of identifying GCs with chaotic dynamics
using the example of the gravitational potential,
which we traditionally use to analyze the orbital
motion of GCs (Bajkova and Bobylev, 2022; Bajkova
et al., 2023a, b). We omit technical details
related to the description of the gravitational potential
model (both axisymmetric and non-axisymmetric),
data, and the selection of GCs. The most detailed
justification and description of the model of the gravitational
potential of the Galaxy, including a three component
axisymmetric part (bulge, disk, halo) and
a built-in central elongated bar, as well as astrometric
data from the Gaia spacecraft, necessary for the formation
of a 6D-phase space for integrating orbits, is
given in the already mentioned works (Bajkova et al.,
2023a, b), where a selection of 45 GSs was made and
all the necessary literary references are given.

The list of GCs is presented in Table~1, where we
also include the results of the analysis obtained within
the framework of this work. The following methods
were used for classification: the method of calculating
the maximum characteristic Lyapunov exponents
(MCLE) (in the classical version and in the version
with renormalization of the shadow orbit), MEGNO
(Mean Exponential Growth factor of Nearby Orbits),
Poincar\'e sections, the frequency method based on
calculating the fundamental frequencies, visual assessment
using images of the reference and shadow
orbits. In addition, a new method is proposed based
on calculating the amplitude spectrum of the orbit
as a function of time and the entropy of the amplitude
spectrum as a measure of orbital chaos (Chumak,
2011). As a model of the bar, we consider
an elongated triaxial ellipsoid with the most probable
parameters known from the literature (e.g., Palous
et al., 1993; Sanders et al., 2019): mass $10^{10} M_\odot$,
length of the major semi-axis of 5 kpc, angle of inclination
to the $X$ axis of 25$^o$, rotation velocity of
40 km s$^{-1}$ kpc$^{-1}$.

In this paper we consider methodological issues
related to the analysis of the regularity/chaoticity of
the orbital motion of GCs based on the methods
listed above, which are discussed in Section 2. In
Section 3, a comparison of the obtained results of the
classification of GCs is carried out. The main results
of the work are formulated in Section 4 (CONCLUSIONS).

\section{METHODS AND RESULTS OF ANALYSIS
OF REGULARITY/CHAOTICITY OF ORBITAL
DYNAMICS OF GLOBULAR CLUSTERS}

\subsection{\it Direct Method for Calculating
MCLE - Probabilistic Method}

We calculate the maximum values of the CLE
(MCLE) using the "shadow" trajectory method according
to the following formula (Melnikov, 2018):

\begin{equation}
\label{Lyap}
L(n)=\frac{1}{n \delta t} \sum_{i=1}^{n} \ln{\frac{D_i}{D_{i-1}}},
\end{equation}
where $D_i$ is the distance in three-dimensional space
between the reference and shadow phase points at
the $i$-th integration step, $D_0$ is the length of the
displacement vector at the initial moment of time, $\delta t$
is the value of the integration step over time. The
true value of the MCLE is equal to the limit of $L(n)$
for $n \rightarrow \infty$ and $D_0 \rightarrow 0$. In practice, the value of the
MCLE is taken to be $L(n)$ obtained for a large value of
$n$. In this case, non-zero positive values of the MCLE
indicate a chaotic motion, while zero and negative
values indicate a regular motion.

In the classical version, that is, in the absence of
renormalization of the shadow orbit, formula (1) is
transformed into a simpler one:

\begin{equation}
\label{Lyapp_1}
L(n)=\frac{1}{n \delta t} \ln{\frac{D_n}{D_0}}.
\end{equation}

The $L(n)$ dependences obtained for the orbits of all
45 globular clusters with the following perturbation of
the initial phase point: $x_1=x_0+x_0\times 0.00001,~y_1=y_0+y_0\times 0.00001,~z_1=z_0+z_0\times 0.00001$, are shown
in Fig.~1. It is evident from the graphs that the approximations
of the Lyapunov exponents are positive
and tend to zero with increasing $n$. In addition, the
set of 45 $L(n)$ functions is divided into two families,
which is especially evident in the graphs constructed
over a shorter time interval of 20 billion years. This
indicates bimodality in the distribution of the MCLE
approximations.

The histogram of the distribution of the MCLE
approximations for t = 12 billion years is shown in
Fig. 2a, where this bimodality is clearly visible. Approximating
the histogram with two Gaussian distributions
using the least squares method and calculating
for each GC the probability of belonging
to one or the other distribution, we obtain a probabilistic
method for dividing the entire set of GCs
into two subsets with relatively small (left Gaussian)
and large (right Gaussian) MCLE values, indicating
different degrees of divergence between the reference
and shadow orbits. As a result of applying the probabilistic
method, we obtained two lists.

The first list included 26 GCs with minimal discrepancies
between the reference and shadow orbits:
NGC6266, Terzan 4, Liller 1, NGC6380,
Terzan 1, Terzan 5, Terzan 6, Terzan 9,
NGC6522, NGC6528, NGC6624, NGC6637,
NGC6717, NGC6723, Terzan 3, NGC6304,
NGC6304, Pismis 26, NGC6569, ESO456-78,
NGC6540, Djorg 2, NGC6171, NGC6316,
NGC6388, NGC6539, NGC6553.

The second list includes 19 GCs with noticeable
discrepancies between the reference and shadow
orbits over long time intervals:
NGC6144, ESO452-11, NGC6273, NGC6293,
NGC6342, NGC6355, Terzan 2, BH229,
NGC6401, Pal 6, NGC6440, NGC6453,
NGC6558, NGC6626, NGC6638, NGC6642,
NGC6256, NGC6325, NGC6652.

In our opinion, the first list includes GCs with regular
orbits, the second list includes those with chaotic
ones.

In order to explain the obtained bimodality of the
MCLE approximations, we constructed the dependences
$f(n)=\ln{\frac{D_n}{D_0}}$
on $n$ for each GC and obtained
two types of graphs, typical for the GC from the first
and second lists, shown in Fig.~3a and 3b, respectively.
The different nature of the obtained dependences
$f(n)$ is visible. The values of the function $f(n)=\ln{\frac{D_n}{D_0}}$
as $n$ tends to infinity saturate to a fairly
large value in both cases. However, in the first case
it is somewhat lower than in the second and has a
smoother and flatter shape. A graphical illustration
for all 45 GCs is given in Fig.~13
in the fifth horizontal row of panels from the top.
Now the nature of the "hyperbolic" dependence $L(n)$
(Fig.~13), calculated using formula (1) and asymptotically
tending to zero with increasing $n$ not only for
regular but also for chaotic orbits, is clear, and the
reason for the bimodality of the distributions of the
MCLE approximations of our sample is also understandable.

Note that here we have considered the case of calculating
the MCLE approximations without renormalizing
the shadow orbit (formula (2)). Naturally,
this algorithm, due to the obtained dependencies
$f(n)=\ln{\frac{D_n}{D_0}}$, when a very large deviation of the
reference and shadow phase points is observed,
cannot be used to correctly calculate the MCLE.
When calculating according to formula (1), a periodic
renormalization of the position of the shadow phase
point relative to the reference one should be carried
out by the distance $D$ between them, so that this
distance is always relatively small. However, the
procedure without renormalization is quite suitable
for separating regular and chaotic orbits due to the
difference in the dependencies $f(n)=\ln{\frac{D_n}{D_0}}$. In
what follows, we will simply call this algorithm probabilistic,
since it separates GCs according to the principle
of maximum probability of belonging to the set of GCs
with regular or chaotic motion.

The designations of the classification of GCs with
regular (R) and chaotic (C) motion, obtained by the
probabilistic method, are given in the third column of
Table~1.

\begin{figure*}
{\begin{center}
              \includegraphics[width=0.7\textwidth,angle=-90]{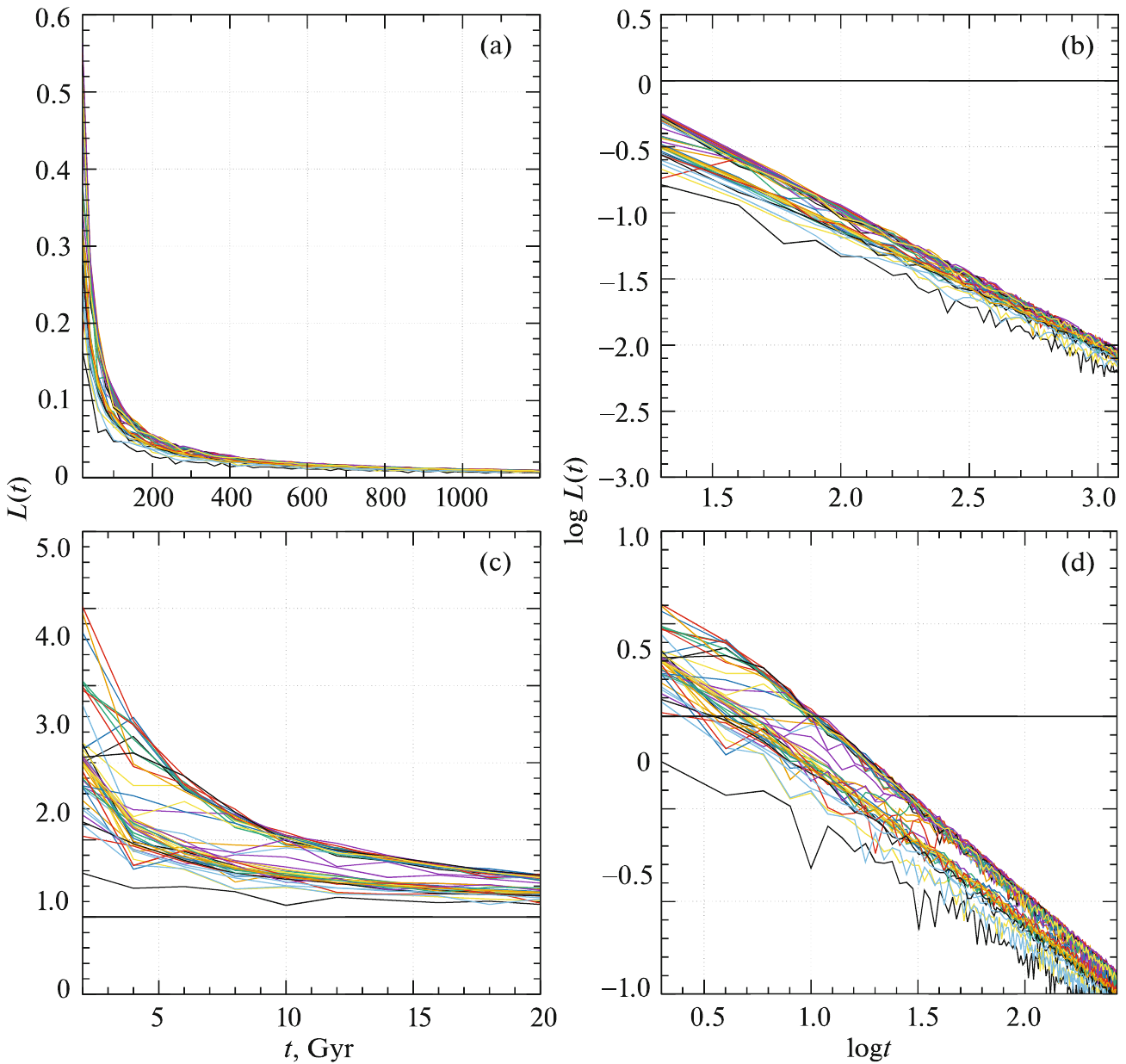}
\caption{\small Calculated approximations of the MCLE without renormalization of the shadow orbit as a function of time for 45 GCs.
Panels (a) and (b) -- the maximum time interval is 1200 billion years, (c) and (d) -- 20 billion years; in panels (a) and (c) the
functions are presented in a linear scale, in (b) and (d) -- in a logarithmic one.}
\label{Lyap1}
\end{center}}
\end{figure*}

\begin{figure*}
{\begin{center}
               \includegraphics[width=0.7\textwidth,angle=-90]{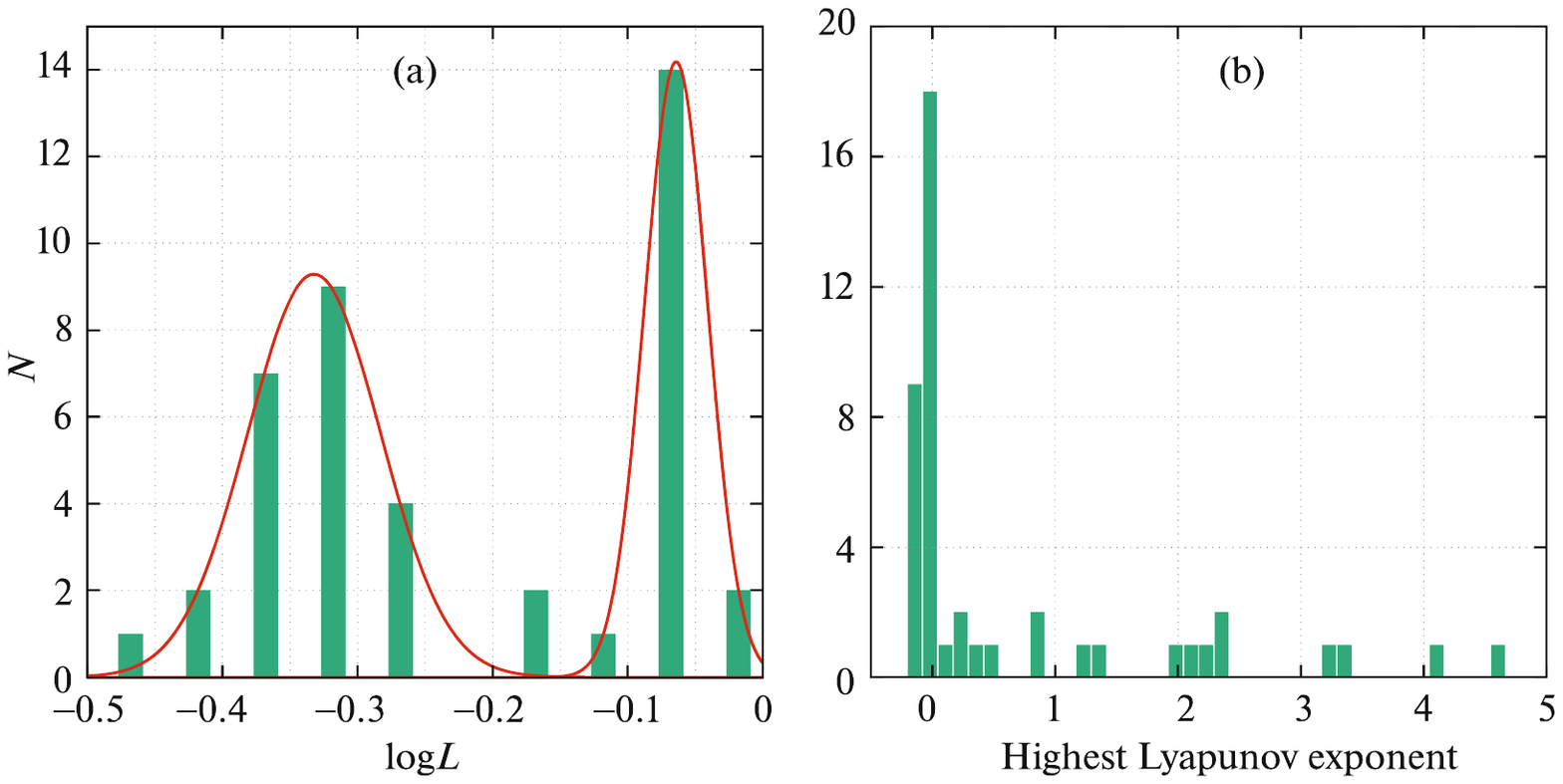}
\caption{\small Histogram of the distribution of MCLE approximations: (a) -- without renormalization of the shadow orbit for 45 GCs
at $t = 12$ billion years; (b) with renormalization of the shadow orbit, on the interval of 120 billion years. Approximation of
the histogram in panel (a) by two Gaussians (red line) allows implementing the probabilistic method of separating GCs with
regular and chaotic orbits (left and right Gaussians, respectively).}
\label{Lyap2}
\end{center}}
\end{figure*}

\begin{figure*}
{\begin{center}
               \includegraphics[width=0.7\textwidth,angle=-90]{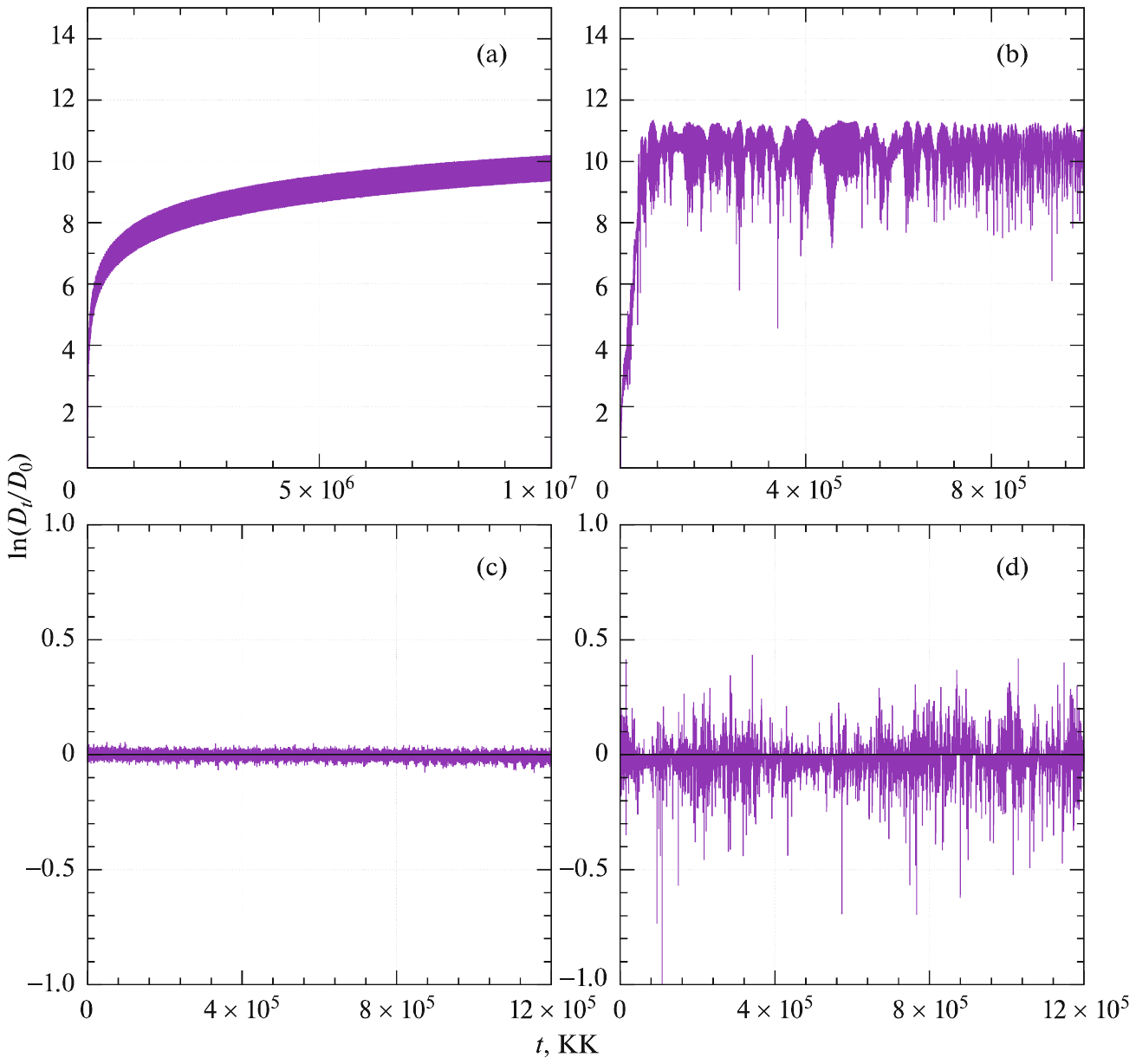}
\caption{\small Typical examples of the $\ln (\frac{D_t}{D_0})$
dependence for regular (a) and chaotic (b) orbits in the case of calculating the MCLE
without renormalizing the shadow orbit. Here, discrete KK samples act as time t, the time step is 0.1 million years, i.e.
$t=0.1\times KK$ million years. Examples are given for NGC6266 (regular orbit) and NGC6355 (chaotic orbit). The $\ln (\frac{D_t}{D_0})$ dependences in the case of renormalizing the shadow orbit are given in panel (c) for NGC6266 and panel (d) for NGC6355. Panels (a) and (b) are shown for a time interval of 100 billion years, panels (c) and (d) -- for 120 billion years.
}
\label{Lyap3}
\end{center}}
\end{figure*}

\subsection{\it Calculation of the MCLE with Renormalization of the Shadow Orbit}

To correctly calculate the MCLE, the vector of the
shift of the shadow phase point from the reference
point was renormalized at each small time interval $\Delta t$
so that the phase point of the shadow orbit shifted
along the vector separating the phase points of the
reference and shadow orbits back to the initial value
of the shift modulus $D_0$ (see, for example, Fig. 9.9
in Murray and Dermott, 1999). If $n_t$ steps are made,
then the MCLE estimate is given by the modified formula
(Murray and Dermott, 1999; Morbidelli, 2014):
\begin{equation}
\label{Lyap_2}
L(t)=\frac{1}{n_t \Delta t} \sum_{i=1}^{n_t} \ln{\frac{D_i}{D_0}},
\end{equation}
where $t=n_t\Delta t$ is the total integration interval.

For the GCs of our sample, we have established
empirically that the optimal size of the renormalization
interval $\delta t$ is a value equal to (30--50)$\Delta t$, where
$\delta t$ is the step of orbit integration over time (see also
formula (1)). In our case $\delta t=0.1$ million years for all
GCs. The value of $\Delta t$ for each GC was selected individually
from the given interval. The obtained dependences
$\ln (\frac{D_t}{D_0})$
on the time interval of 120 billion
years in the case of shadow orbit renormalization
are given in the bottom row of panels of Fig.~3 for
NGC6266 (c) and for NGC6355 (d). Comparison
with similar dependences obtained without renormalization
of the shadow orbit (Figs. 3a and 3b) shows
that renormalization led to a significant decrease in
the distance between the phase points of the reference
and shadow orbits. This allowed us to correctly calculate
the MCLE approximations, the values of which
in the case of renormalization on the selected time
interval of 120 billion years were -0.017 (less than
zero) for NGC6266 and 2.257 (greater than zero) for
NGC6355, which indicates the regularity of the orbit
in the first case and chaos in the second.

We calculated the MCLE approximations with
shadow orbit renormalization for all 45 GCs in our
sample. The corresponding histogram of the MCLE
value distribution is shown in Fig.~2b. The list of
GCs with regular orbits includes objects with the
approximation values MCLE$<0$, while the list of
GCs with chaotic orbits includes objects with the
values MCLE$>0$. The MCLE approximation values
together with the orbit designations -- regular (R) or
chaotic (C) -- are given in the fourth column of Table~1. Note that this classification differs somewhat
from the previous one, obtained by the probabilistic
method (see also the third column of Table~1). The
correlation coefficient between the results of orbit
classification by these two methods (I and III) is
$K_c = 0.643$ (see Table~2). A detailed comparison
of the results of the analysis of the regularity of the
GC orbits obtained by different methods will be given
below, in Section 3.

\subsection{\it MEGNO}

A description of MEGNO (Mean Exponential
Growth factor of Nearby Orbits) can be found in
Morbidelli (2014) and Melnikov (2018). MEGNO
is one of the most widely used methods for detecting
chaos in various problems of celestial mechanics, and
its application is possible at significantly shorter times
compared to MCLE. When analyzing the regularity
of orbits, we use the property of MEGNO that in the
case of a regular trajectory $M(t)\rightarrow 2$ as $t \rightarrow \infty$.

\begin{figure*}
{\begin{center}

              \includegraphics[width=0.75\textwidth,angle=-90]{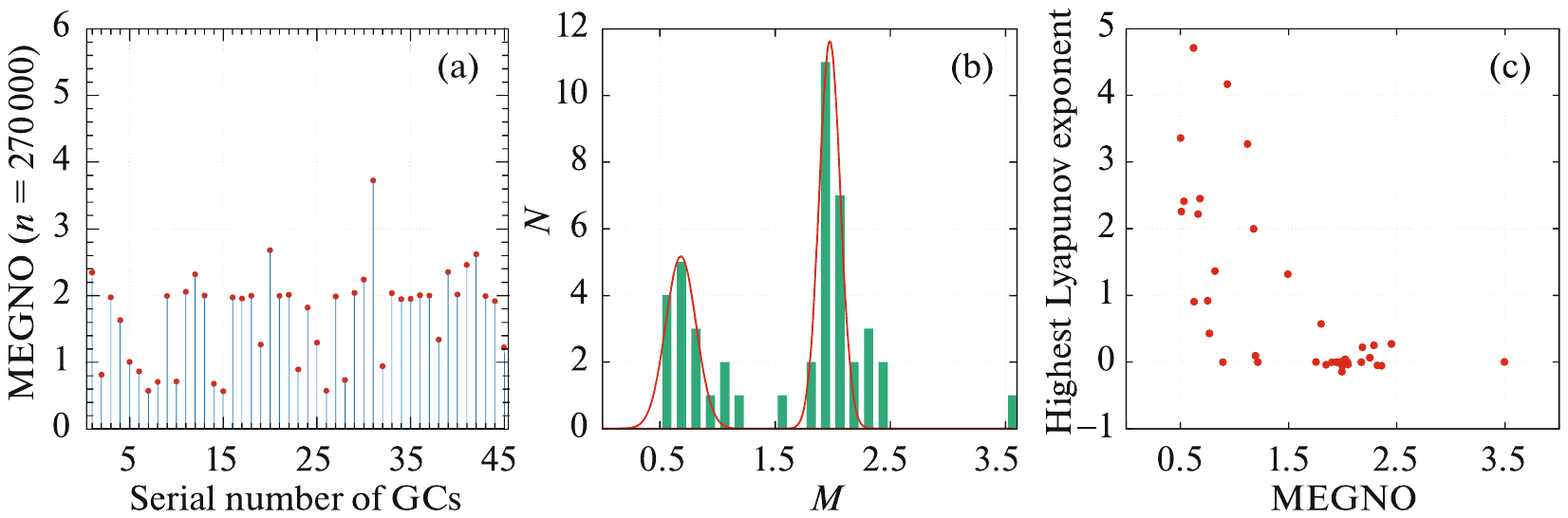}
\caption{\small MEGNO approximations for 45 GCs on a time interval of 270 billion years (a). Histogram of the distribution of
MEGNO approximations and its approximation by two Gaussians (red line), allowing to implement a probabilistic method for
separating GCs with regular and chaotic orbits (b). "MEGNO -- MCLE" diagram (c).}
\label{Lyap4}
\end{center}}
\end{figure*}

In this paper, we use the method well described by
Melnikov (2018) and proposed in Breiter et al. (2005)
for the case of numerical integration with a constant
step, which is as follows: the value of the MEGNO
parameter at the n-th integration step is determined
by the formula
\begin{equation}
\label{megno}
m(n)=\frac{n-1}{n} m(n-1) + 2 \ln{\frac{D_n}{D_{n-1}}}.
\end{equation}
For the time-averaged value of MEGNO we have:
\begin{equation}
\label{megno1}
M(n)=\frac{1}{n} ((n-1)M(n-1) + m(n)),
\end{equation}
assuming $m(0)=0$ and $M(0)=0$.

For the correct calculation of MEGNO values,
as well as for the correct calculation of MCLE values,
it is necessary to renormalize the position of the
shadow phase point relative to the reference point
by the distance $D$ between them. But since estimates
of the MCLE value using MEGNO are characterized
by less reliability than direct calculation of
MCLE (see Melnikov, 2018 and references therein),
we limited ourselves to calculating MEGNO without
renormalizing the shadow orbit in order to obtain
only a qualitative result suitable for separating regular
and chaotic orbits similar to the probabilistic method
based on calculating MCLE without renormalization
(see Section 2.1).

The result of applying formulas (4) and (5) to
our sample over a time interval of 270 billion years
($n = 270 000$, $\delta t=1$ Myr) is shown in Fig.~4. The
values of the MEGNO approximations for 45 GCs are
shown in panel (a) (the abscissa axis shows the serial
numbers of the GCs in accordance with Table~1).
The histogram of the distribution of the MEGNO
approximations and its approximation by two Gaussians
(red line), which allows implementing the probabilistic method
of separating GCs with regular (right
Gaussian centered at point $M\approx2$) and chaotic (left
Gaussian) orbits, are shown in panel (b).

The lists of GCs with regular and chaotic orbits
obtained on the basis of MEGNO and MCLE without
shadow orbit normalization practically coincide (with
the exception of NGC6144 and NGC6440) (see
the fourth column of Table~1, where the values of
MEGNO approximations for each GC and the orbit
classification designations are given -- (R) or (C)).
The correlation coefficient between these classification
results is $K_c=$0.911 (see Table 2).

The "MEGNO -- MCLE" with shadow orbit renormalization
diagram is shown in Fig.~4c. The correlation
coefficient between the calculated values of the
MEGNO and MCLE approximations is $K_c=$0.70.

\subsection{\it Poincar\'e Sections}

One of the methods for determining the nature
of the motion (regular or chaotic) is the analysis of
Poincar\'e sections (Murray and Dermott, 1999). The
algorithm we used to construct the mappings is as
follows:

1) we consider the phase space $(X,Y,V_x,V_y)$;

2) we exclude $V_ Y$ using the conservation law of the
Jacobi integral and move to the space $(X,Y,V_x)$;

3) we define the plane $Y=0$, we designate the
points of intersection with the orbit on the plane
$(X,V_x)$; we take only those points at which
$V_y>0$.

Similarly, the phase space $(Y,Z,V_y,V_z)$ or
$(R,Z,V_R,V_z)$ can be considered. Then the Poincar\'e
sections will be reflected on the plane $(Y,V_y)$ or
$(R,V_R)$, respectively.

If the intersection points of the plane add up to a
continuous smooth line (or several separated lines),
then the motion is said to be regular. In the case of
chaotic motion, instead of being located on a smooth
curve, the points fill a two-dimensional region of
phase space, sometimes with the effect of sticking
points to the boundaries of islands corresponding to
ordered motion (Murray and Dermott, 1999).

Figure~5 shows an example of Poincar\'e sections
for regular (NGC6266) and chaotic (NGC6355)
motion.

\begin{figure*}
{\begin{center}
                \includegraphics[width=0.7\textwidth,angle=-90]{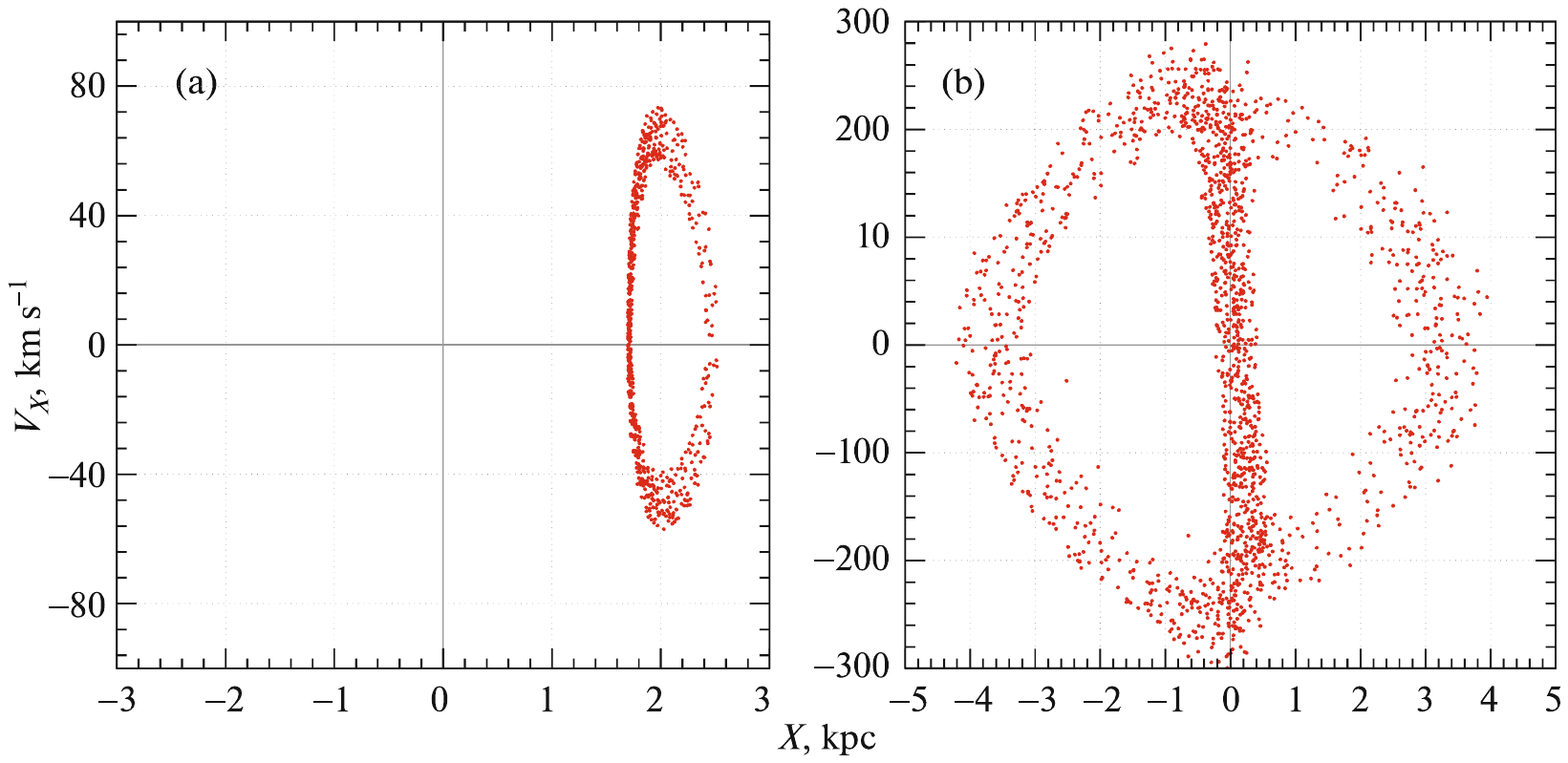}
\caption{\small Poincar\'e sections: (a) for NGC6266 (regular orbit); (b) NGC6355 (chaotic orbit).}
\label{Lyap5}
\end{center}}
\end{figure*}

We calculated the Poincar\'e sections $(X,V_x)$ for
all 45 GCs and visually determined the nature of their
motion -- (R) or (C). The classification results are reflected
in the sixth column of Table~1, and the graphical
representation of the sections in
Fig.~13 in the fourth (from the top) horizontal row of
panels.

\subsection{\it Frequency Method}

Another way to study the regularity/randomness
of orbits is to use orbital frequencies. Valluri et al.
(2010) and Nieuwmunster et al. (2024) (see Section
3.1 in the latter paper) showed that it is possible to
measure the stochasticity of an orbit based on the
shift of fundamental frequencies determined in two
consecutive time intervals. For each frequency component
$f_i$, a parameter called the frequency drift is
calculated:
\begin{equation}
\label{freq}
\log(\Delta f_i)=\log|\frac{\Omega_i(t_1)-\Omega_i(t_2)}{\Omega_i(t_1)}|,
\end{equation}
where i defines the frequency component in Cartesian
coordinates (i.e. $\log(\Delta f_x), \log(\Delta f_y)$ and $\log(\Delta f_z)$).
Then the largest value of these three frequency drift
parameters $\log(\Delta f_x)$ is assigned to the frequency drift
parameter $\log(\Delta f)$. The higher the value of $\log(\Delta f)$,
the more chaotic the orbit. However, as shown by
Valluri et al. (2010), the accuracy of frequency analysis
requires at least 20 oscillation periods to avoid
classification errors.

\begin{figure*}
{\begin{center}
               \includegraphics[width=0.7\textwidth,angle=-90]{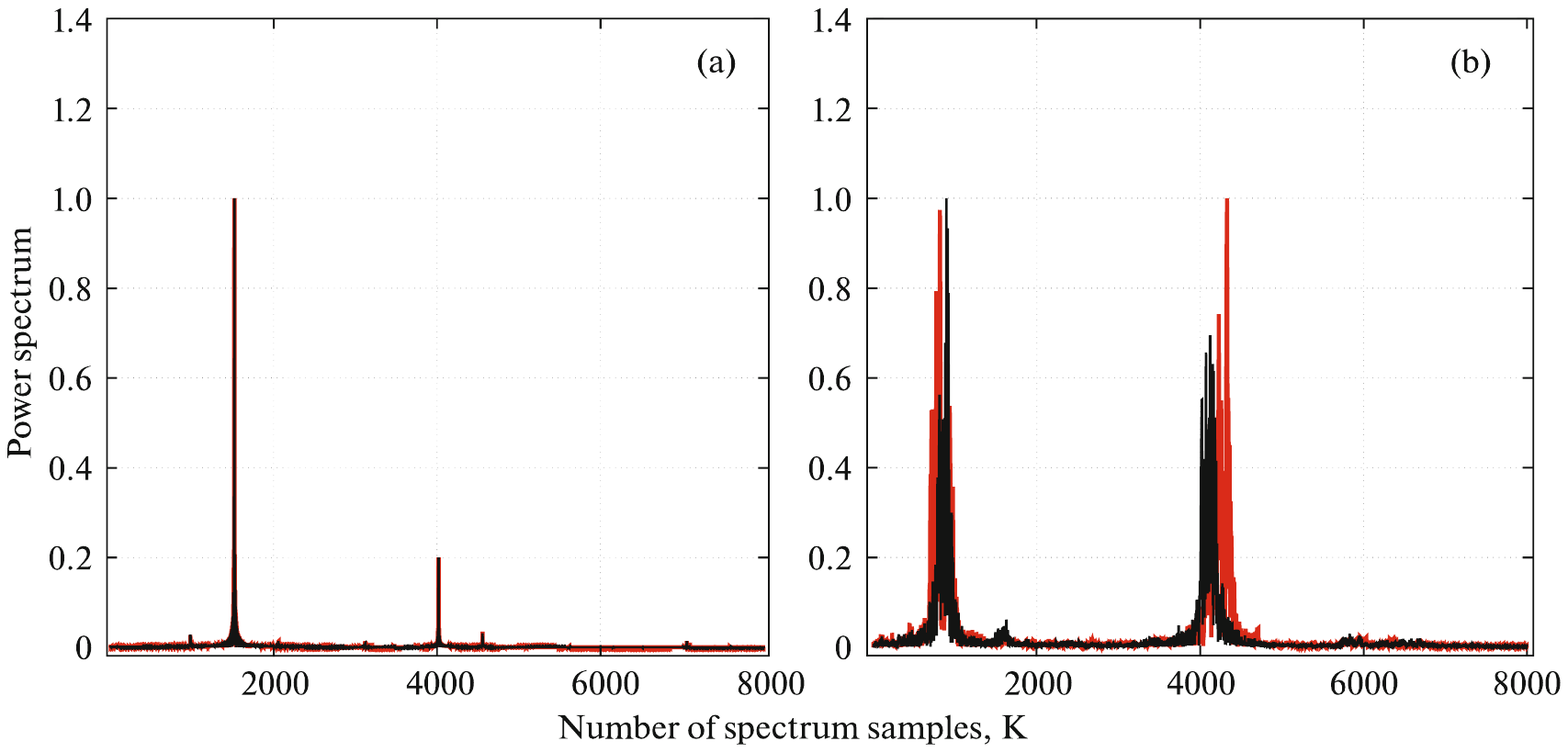}
\caption{\small Illustration of the frequency method, the amplitude spectrum of the first half of the time sequence is shown in red, the second half in black. Panel (a) refers to NGC6266 with a regular orbit, panel (b) to NGC6355 with a chaotic orbit.}
\label{Lyap6}
\end{center}}
\end{figure*}

We determined the frequency drift parameter
$\log(\Delta f)$ for all 45 GCs, which was used to establish
the nature of their motion -- (R) or (C). The series
$x(t_n), y(t_n), z(t_n)$ were determined on the time
interval [0, 120] billion years. The first amplitude
spectrum of each GC was calculated on the time
interval [0, 60] billion years, the second one on [60, 120] billion years. Then the frequency drift parameters
for each time series $x(t_n), y(t_n), z(t_n)$ were
determined using formula (6), and the largest value
of them was taken as the frequency drift parameter
$\log( \Delta f)$. In the case of the coincidence of fundamental
frequencies $\Omega_i(t_1)=\Omega_i(t_2)$, we artificially assumed
the frequency drift parameter to be equal to $-4$.

The results of calculating the frequency drift parameter
and the nature of the GC motion -- (R) or (C), which was determined from a joint analysis of the
frequency drift parameter values and a visual analysis
of the amplitude spectra, are reflected in the seventh
column of Table 1 (see also Fig.~13, sixth
row). It turned out that the threshold value for separating
regular and chaotic orbits is $\log(\Delta f)=-2.14$.
Regular orbits correspond to values smaller or equal
to the threshold, and chaotic orbits to values larger
than the threshold.

Figure~6 shows an example of spectra for regular
(NGC6266) and chaotic (NGC6355) motion, where
the amplitude spectrum of the first half of the time
sequence is shown in red, and the second half in
black. Fig.~7a shows a histogram of the frequency
drift parameter distribution $\log(\Delta f)$, panel (b) shows
a "frequency drift -- MCLE" diagram showing a good
correlation between the values of the frequency drift
parameter and MCLE, the correlation coefficient is
$K_c=0.76$. A graphical illustration
of the method for all 45 GCs is shown in Fig.~13 in the
sixth horizontal row of panels from the top. The correlation
coefficients of the classification results with
other methods are given in Table~2. The lowest correlation
($K_c = 0.687$) is observed with the results of the
MEGNO, the highest
(0.910) -- with the Poincar\'e section method.

\begin{figure*}
{\begin{center}
               \includegraphics[width=0.7\textwidth,angle=-90]{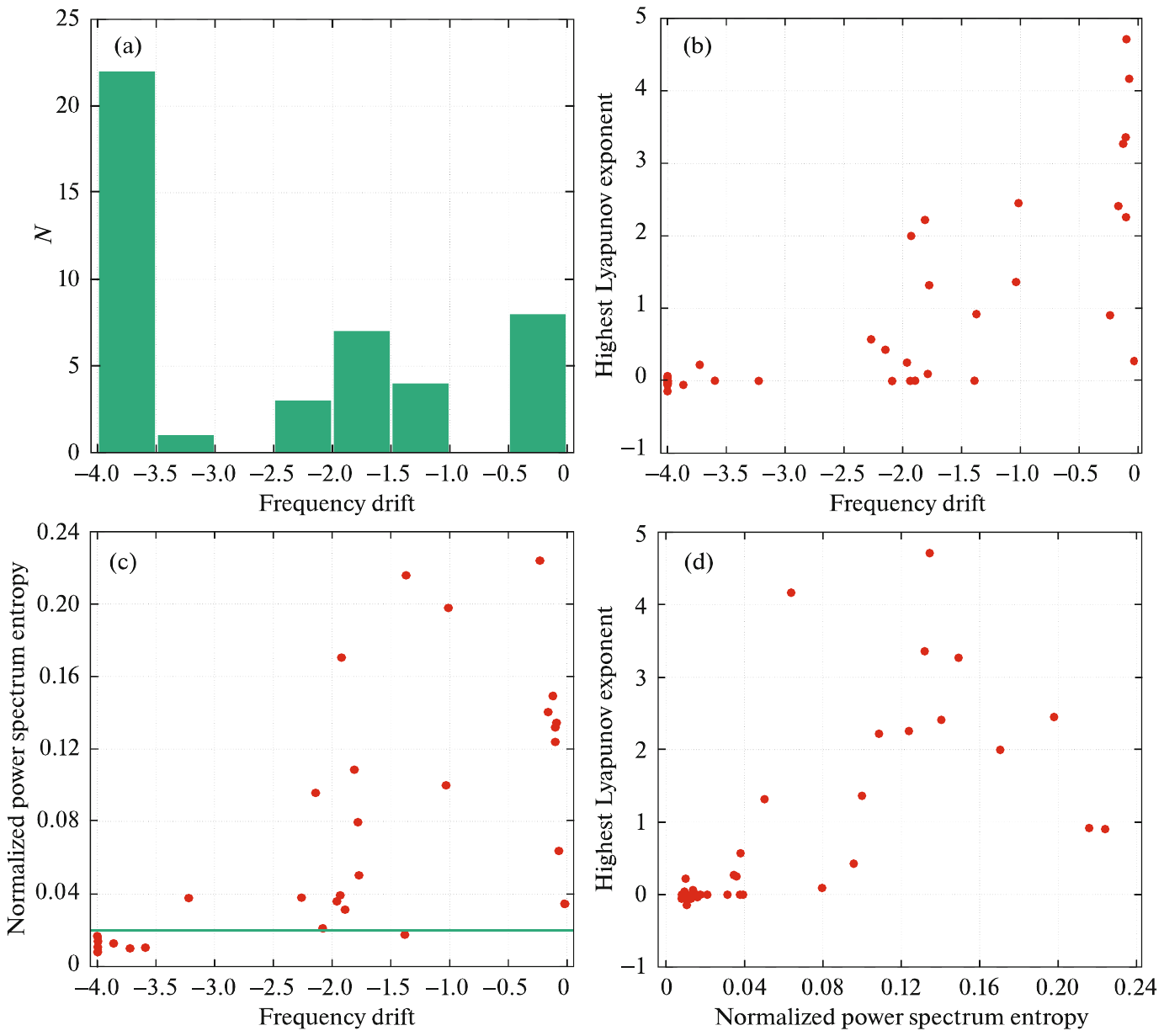}
\caption{\small Histogram of the frequency drift parameter distribution $\log(\Delta f)$ (a); "frequency drift -- MCLE" diagram (b); "frequency drift -- entropy of the normalized amplitude spectrum of the orbit" diagram, the threshold entropy value is marked by the green line (c); "entropy of the normalized amplitude spectrum of the orbit -- MCLE" diagram (d).}
\label{Lyap6}
\end{center}}
\end{figure*}

\subsection{\it Visual Assessment of Orbital Regularity}

The most informative illustration of the discrepancy
between the reference and shadow phase points
is Fig.~13, which shows the reference
and shadow orbits for each GC in the order (from
left to right) as they are given in Table~1. The uppermost
panels show the radial values of the orbit
as a function of time over the interval $[0,-12]$ billion
years, comparable with the age of both the GC
and the Universe; below, vertically (the second and
third horizontal rows), are the projections of the orbits
$X-Y$, $X-Z$, $Z-Y$, respectively, constructed in the rotating
bar system over the time interval $[-11,-12]$ billion
years. In all the graphs, the reference orbits are
shown in yellow, and the shadow orbits in purple.
Many objects in the graphs have only purple color,
which means that the shadow orbit almost completely
coincides with the reference orbit (yellow lines are
covered with purple). Such objects include GCs with
regular orbits. In the graphs of GCs with chaotic
orbits, both purple and yellow lines are visible, which
allows us to qualitatively judge the degree of orbital
chaos. The results of the visual assessment -- (R)
or (C) -- are given in the eighth column of Table 1.
Figure 8 shows an illustration for the GC NGC6266
with a regular orbit and NGC6355 with a chaotic
orbit as an example.

\begin{figure*}
{\begin{center}
               \includegraphics[width=0.7\textwidth,angle=-90]{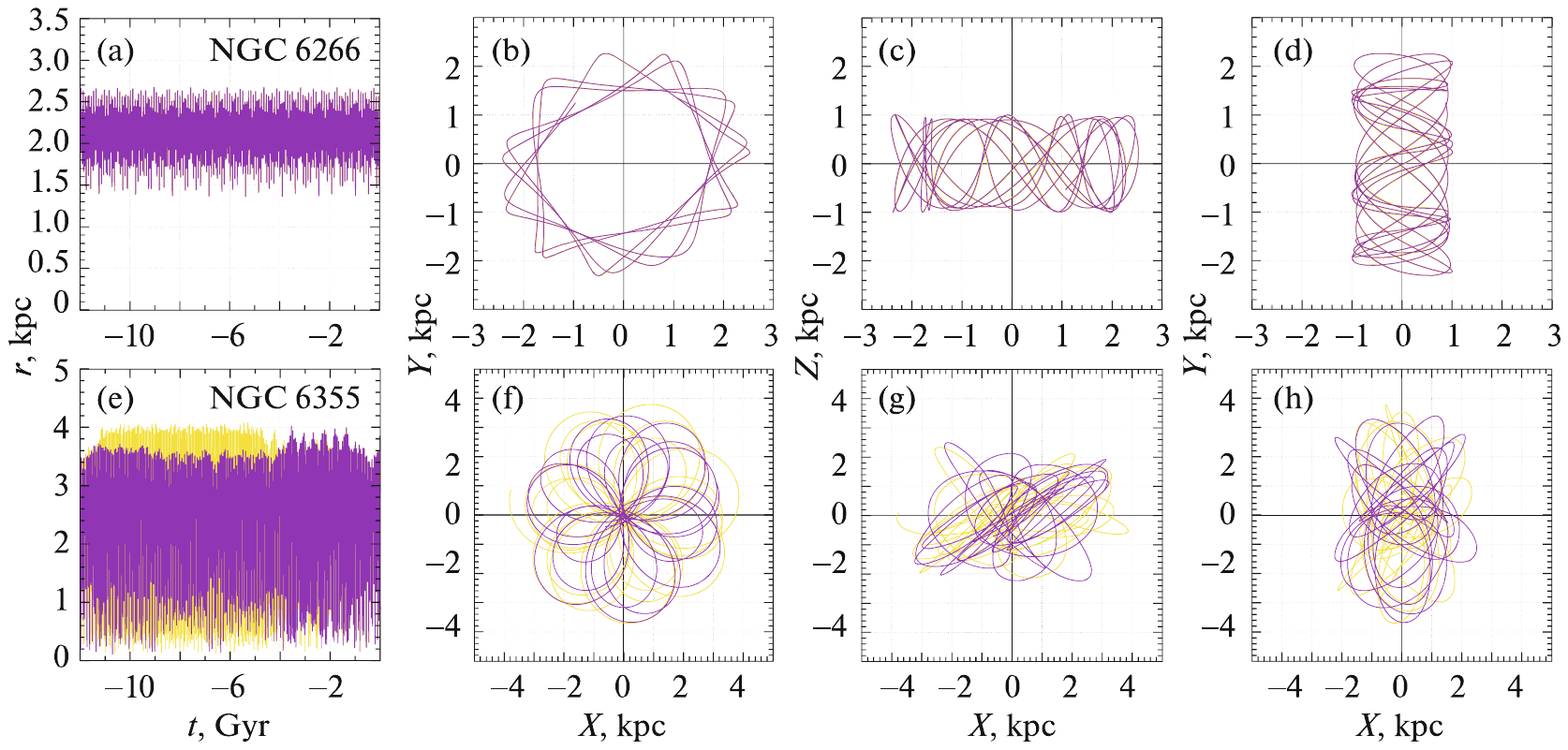}
\caption{\small Orbits of globular clusters NGC6266 (R) (a)--(d), NGC6355 (C) (e)--(h). In the panels from left to right: radial values of the orbit depending on time on the interval $[-12, 0]$ billion years; $X-Y$, $X-Z$, $Z-Y$ projections of orbits constructed in the rotating bar system on the time interval $[-12, -11]$ billion years (the reference orbit is shown in yellow, the shadow orbit is
shown in purple).}
\label{Lyap7}
\end{center}}
\end{figure*}

\subsection{\it New Method Based on Spectral Analysis}
In this case, the spectral analysis of the orbits
is based on the calculation of the modulus of the
discrete Fourier transform (DFT) of the uniform time
series of radial distances of the orbital points from
the center of the Galaxy $r_n$, calculated from their
$X, Y, Z$ galactic coordinates $X(t_n), Y(t_n), Z(t_n)$ as
functions of time: $$r(t_n)=\sqrt{X(t_n)^2+Y(t_n)^2+Z(t_n)^2}$$,
where $n=0,...,N-1$ ($N$ is the series length).

Thus, the formula for the DFT modulus (amplitude
spectrum) of the $r_n$ sequence will look like this:
\begin{equation}
\label{Equ1}
R_k=|\frac{1}{N} \sum_{n=0}^{N-1} r_n\exp{(-\jmath\frac{2\pi\times n\times k}{N})}|,
\end{equation}
where $k=0,..., N-1$. The length of the series is
chosen to be $N=2^\alpha$, where $\alpha$ is a positive integer
value, so that the fast Fourier transform algorithm can
be used to calculate the DFT. The required length of
the series is achieved by supplementing the real series
with zeros.

In our case, the length of the real sequences is
120 000, since we integrate the orbits 120 billion years
ago with an integration interval of 1 million years.
Before calculating the DFT, we pre-center the coordinate
series (i.e., get rid of the constant component),
then supplement the resulting sequence $r_n$ with zero
readings at $n>120000$ until the length of the entire
analyzed sequence reaches $N=262144=2^{18}$. Note
that supplementing the original sequence with zeros
is also useful from the point of view of increasing
the accuracy of the coordinates of the spectral components.
Since the interval between the sequence
readings in time is $\Delta_t=0.001$ billion years, the analyzed
frequency range, which is a periodic function,
is $F=1/\Delta_t=1000$  billion years$^{-1}$. The frequency
discrepancy is $\Delta_F= F/N \approx 0.03815$ billion years$^{-1}$.
For convenience, in what follows, we will indicate not
physical frequencies in the graphs, but the numbers
of readings $k$ (or $K$) of the discrete Fourier transform
(formula (7)). One can go from $k$ to the physical
frequency using the formula $f=k\times\Delta_F\approx k\times 0.003815$. Then, the obtained amplitude spectrum
of the GC orbit is normalized so that the maximum
value is equal to unity.

The decision on the nature of the orbital dynamics
of the GC is determined by calculating the Shannon
entropy of the normalized amplitude spectrum $R_k$ as
a measure of chaos (Chumak, 2011):
\begin{equation}
\label{Equ2}
E_R=-\frac{1}{M}\sum_{k=0}^{N-1} R_k \ln(R_k),
\end{equation}
where $M$ is a scale factor that is introduced for the
convenience of presenting numerical results.

\begin{figure*}
{\begin{center}
    \includegraphics[width=0.7\textwidth,angle=-90]{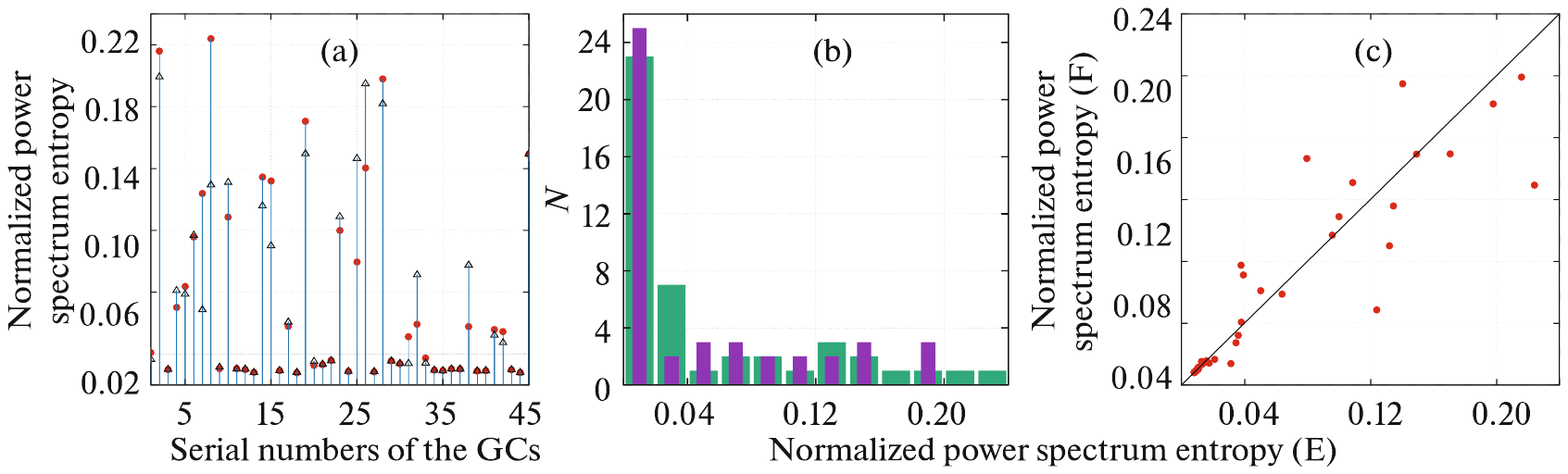}
\caption{\small (Panel (a): Entropy values of normalized amplitude spectra of 45 GCs (ordinal number on the abscissa axis) for the
reference (red dots) and shadow (black triangles) orbits. Panel (b): Histograms for the entropy of normalized amplitude spectra
of the reference (green) and shadow (purple) orbits of 45 GCs. Panel (c): Diagram comparing the entropy values of normalized
amplitude spectra of the reference (E) (abscissa axis) and shadow (F) (ordinate axis) orbits. }
 \label{fentr}
\end{center}}
\end{figure*}

Obviously, the higher the entropy value, the higher
the degree of orbital chaos. Since we analyze both
reference orbits and shadow orbits obtained by perturbing
the initial phase point, we should also pay
attention not only to the entropy value, but also to the
difference in the entropy values of the reference and
perturbed orbit spectra. In the case of a regular orbit,
this difference should be small enough by analogy
with the Lyapunov exponents. In addition, we note
that by analogy with the Lyapunov exponents, we
take the orbital integration time to be large enough.
In our case, as in the case of the frequency method,
it is 120 billion years, which, as already mentioned, is
almost an order of magnitude greater than the age of
the Universe.

The proposed method was applied to both reference
and shadow orbits with perturbation of the phase
initial point, similar to the algorithm implemented in
Section 2.1.

In Fig.~13 the bottom row shows
normalized amplitude spectra of radial values of reference
and shadow orbits as functions of time on
the interval $[-120, 0]$ billion years, shown in black
and red, respectively. We note immediately that they
have the character of line spectra for GCs with regular
dynamics and wide spectra for GCs with chaotic
dynamics, which leads to a noticeable increase in
entropy in the latter case.

The results of applying the new method are also
clearly demonstrated in Fig.~9.

As can be seen in Fig.~9a, globular clusters with
entropy values less than 0.02 have practically equal
values for the reference and shadow orbits, i.e. the
perturbation of the initial point almost did not lead to
a change in entropy. This also follows from Fig.~9c.
With an increase in entropy, the difference between
the entropy values for the reference and shadow orbits
increases. Therefore, globular clusters with an
entropy value less than 0.02, which is taken as the
threshold for the adopted conditions of orbit integration
and the values of the scale factor $M$ (marked
in green in Fig.~7c), we attribute to globular clusters
with regular dynamics (R), and the rest to GCs
with chaotic dynamics (C). As follows from the histogram
of the entropy distribution for the reference
and shadow orbits (Fig.~9b), as a result of the perturbation
of the initial point, the entropy value for some
globular clusters decreased and became less than
0.02, but when deciding on the nature of the dynamics,
we also look at the entropy increment. For GCs
with regular dynamics, it should be sufficiently small
(as a rule, it is an order of magnitude or more lower
than for GCs with chaotic dynamics). In essence,
the above represents an algorithm for deciding on the
dynamic nature of GCs -- (R) or (C) -- as a result of
applying the proposed method based on the spectral
analysis of the radial values of the orbits as a function
of time over an interval of 120 billion years and calculating
the entropy of the obtained normalized spectra
using the scale factor $M=10 000$. As additional
modeling has shown, with a reasonable change in
the integration time and the scale factor $M$, only
the threshold value of entropy changes, which has
virtually no effect on the classification results. We
consider the parameters adopted in this work to be
optimal.

Based on the results of applying our method, we
assigned the following GCs to the first group of
objects with regular dynamics (R):
NGC6266, Terzan4, Liller1, NGC6380,
Terzan1, Terzan5, Terzan6, Terzan9,
NGC6522, NGC6528, NGC6624, NGC6637,
NGC6717, NGC6723, NGC6304, Pismis26,
NGC6569, ESO456-78, NGC6540, Djorg2,
NGC6171, NGC6539, NGC6553.

The second group of objects with chaotic dynamics
(C) -- the globular clusters
NGC6144, ESO452-11, NGC6273, NGC6293,
NGC6342, NGC6355, Terzan2, BH229,
NGC6401, Pal6, NGC6440, NGC6453,
NGC6558, NGC6626, NGC6638, NGC6642,
Terzan3, NGC6256, NGC6325, NGC6316,
NGC6388, NGC6652.

The first group has 23 GCs, the second -- 22 GCs.

\section{COMPARISON OF CLASSIFICATION RESULTS FOR GLOBULAR CLUSTERS
WITH REGULAR AND CHAOTIC DYNAMICS}
The results of the classification of 45 GCs by
the orbit regularity/chaoticity feature, obtained in
the previous section by seven different methods, are
summarized in Table~1. The determination of the
probability of the final solution $P(R)$ by the new
method based on the calculation of the entropy of
the normalized amplitude spectrum of the orbit as a
function of time is considered the seventh method. A
graphical illustration of the methods of analyzing the
orbit regularity as applied to each GC is given in Fig.~13.

In order to compare the classification results of the
GCs based on the data in Table~1, a correlation matrix
was calculated based on the classification results
obtained by seven methods, when the solution (R)
is assigned the number "1", and the solution (C) is
assigned the number "0". The obtained correlation
matrix is presented in Table~2.

As a result, we can
conclude that the lowest correlation with other methods
is observed with MCLE, and the highest with the
visual method. The Poincar\'e sections method for decision
making is close to the frequency method, both
of these methods correlate well with each other and
with other methods. The correlation of the MCLE,
frequency method and the new method is also well
reflected in the diagrams in Fig.~7.

In general, we can conclude that all methods correlate
well with each other. As a final decision on the
classification of GCs with regular (R) and chaotic (C)
dynamics, we will accept the results of the classification
based on the "voting" principle. As a result, the
list of GCs with regular dynamics included 24 objects:
NGC6266, Terzan4, Liller1, NGC6380,
Terzan1, Terzan5, Terzan6, Terzan9,
NGC6522, NGC6528, NGC6624, NGC6637,
NGC6717, NGC6723, Terzan3, Pismis26,
NGC6569, ESO456-78, NGC6540, Djorg2,
NGC6171, NGC6316, NGC6539, NGC6553.

The list with chaotic dynamics consists of 21 objects:
NGC6144, ESO452-11, NGC6273, NGC6293,
NGC6342, NGC6355, Terzan2, BH229,
NGC6401, Pal6, NGC6440, NGC6453,
NGC6558, NGC6626, NGC6638, NGC6642,
NGC6256, NGC6304, NGC6325, NGC6388,
NGC6652.

Note that the final lists differ from the initial ones
obtained by the probabilistic method (see Section 2.1)
by two objects: NGC6304 and NGC6388, which
migrated from the first list of objects with regular
motion to the list of objects with chaotic motion.
Note that the new method yielded a fairly high
correlation ($K_c=0.825$) with the frequency method,
the Poincar\'e section method and the visual method.
At the same time, the GCs Terzan3, NGC6304,
NGC6316 changed their classification to the opposite.
We believe that the GCs Terzan3 and
NGC6316 can be classified as weakly chaotic. But
this classification issue requires a separate study,
which is planned for the future.

In order to determine the cause of the chaotization
of the GC motion, we also compared such parameters
as the apocentric distance, pericentric distance, and
eccentricity of the reference and shadow orbits for
all 45 GCs 12 Gyr ago (Fig.~10). Obviously, GCs
with regular orbits should have deviations from the
diagonal comparison line close to zero, which also
follows from the visual analysis of the orbits. As can
be seen from Fig.~10, GCs with chaotic orbits already
give noticeable deviations from the comparison line,
especially in the region of small values of the pericentric
distance (less than 0.4 kpc) and large values
of eccentricity (more than 0.8) due to a stronger
influence of the central bar on the orbital motion of
GCs precisely in the region closest to the center of the
Galaxy. The apocentric distances deviate maximally
from the comparison line in the region of 2.5 -- 3.5 kpc.
The degree of deviations of the orbital parameters is
especially clearly seen in Fig.~11, which shows the
relative deviations of the parameters of the shadow
orbit from the reference orbit. It should be noted that
the relative deviations of the apocentric distances are
small. The largest relative deviations are experienced
by the pericentric distances. Note that the correlation
between the deviations of the pericentric distances
and eccentricities from the corresponding comparison
lines is quite large and amounts to 0.85. These two
features can serve as specific indicators of the chaos of
GC orbits. Thus, GSs with pericentric distances less
than 0.4 kpc and relative deviations of the pericentric
distances $|Peri_E/Peri_F-1|>0.01$, where the index
E refers to the reference orbit, and the index F -- to the
shadow one, are
ESO452-11, NGC6273, NGC6355, Terzan2,
BH229, NGC6401, Pal6, NGC6453,
NGC6558, NGC6626, NGC6638, NGC6642,
NGC6256, NGC6652,
and GCs with eccentricities greater than 0.8 --
ESO452-11, NGC6273, NGC6293, NGC6355,
Terzan2, BH229, NGC6401, Pal6,
NGC6440, NGC6453, NGC6558, NGC6626,
NGC6638, NGC6642, NGC6652.
The intersection of these two sets is 13 objects. All
listed GCs are included in our final list of GCs with
chaotic dynamics.

Thus, we can conclude that the list of GCs with
chaotic dynamics includes mainly globular clusters
with elongated radial orbits, small pericentric distances
(less than 0.4 kpc) and large eccentricities
(more than 0.8). In general, our result is consistent
with the conclusions of Machado and Manos (2016)
that chaotic orbits mainly occupy the bar region.

Figure~12 shows the distributions of the total energy
at the initial moment of time depending on the
serial number of the GC (a) and the diagram "component
of angular momentum $L_z$ -- total energy" (b).
It is evident that the energy distributions are almost
the same for regular and chaotic orbits. However,
GCs with regular orbits have higher values of the
moment $L_z$.

\begin{figure*}
{\begin{center}
               \includegraphics[width=0.7\textwidth,angle=-90]{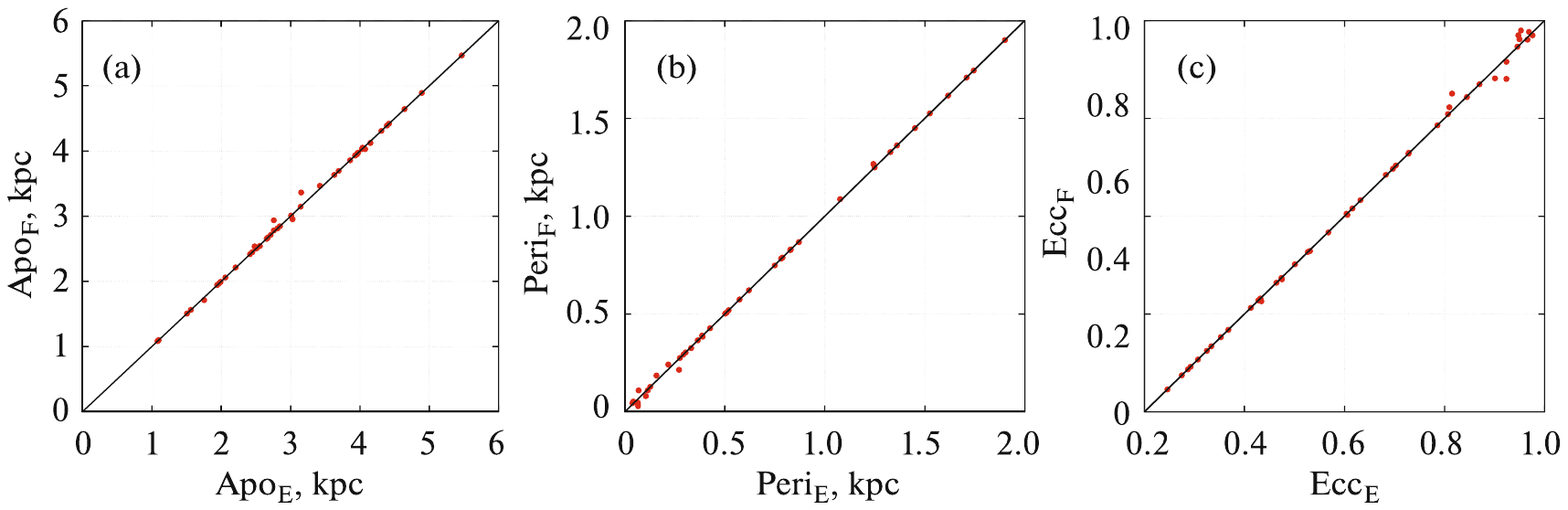}
\caption{\small Comparison of the parameters of the reference (E) and shadow (F) orbits for 45 GCs (apocentric distance (a),
pericentric distance (b), orbital eccentricity (c)).}
\label{Lyap71}
\end{center}}
\end{figure*}

\begin{figure*}
{\begin{center}
               \includegraphics[width=0.7\textwidth,angle=-90]{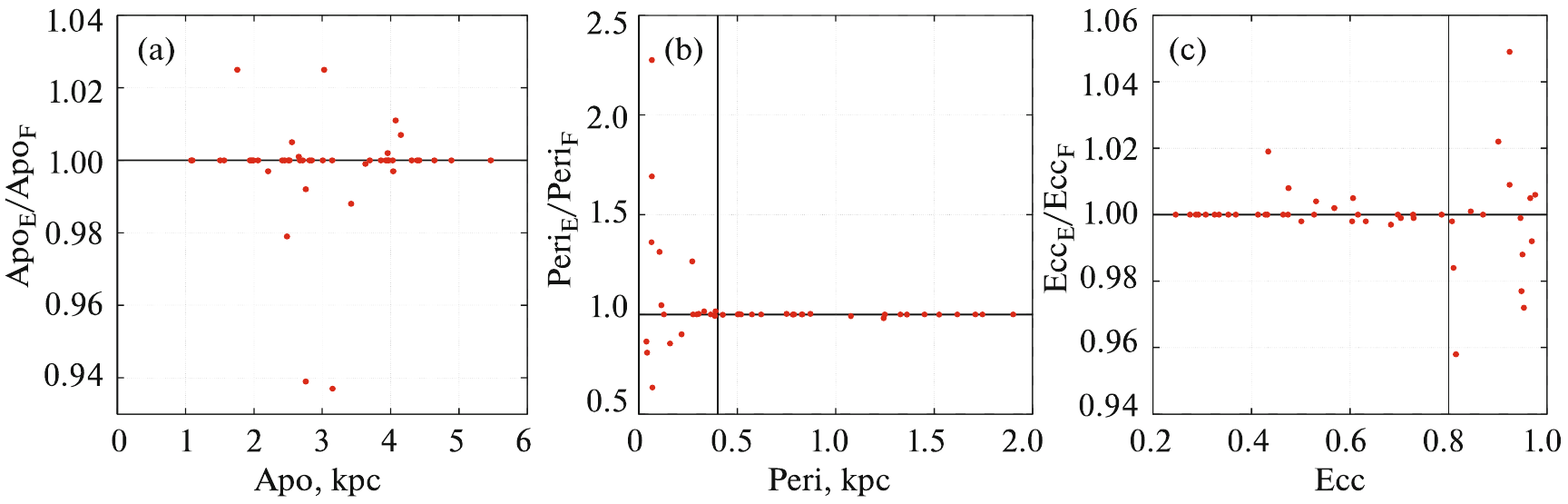}
\caption{\small Relative deviations of the parameters of the shadow(F) fromthe reference (E) orbits of 45GCs (apocentric distance (a),
pericentric distance (b), orbital eccentricity (c)).}
\label{Lyap72}
\end{center}}
\end{figure*}

\begin{figure*}
{\begin{center}
               \includegraphics[width=0.7\textwidth,angle=-90]{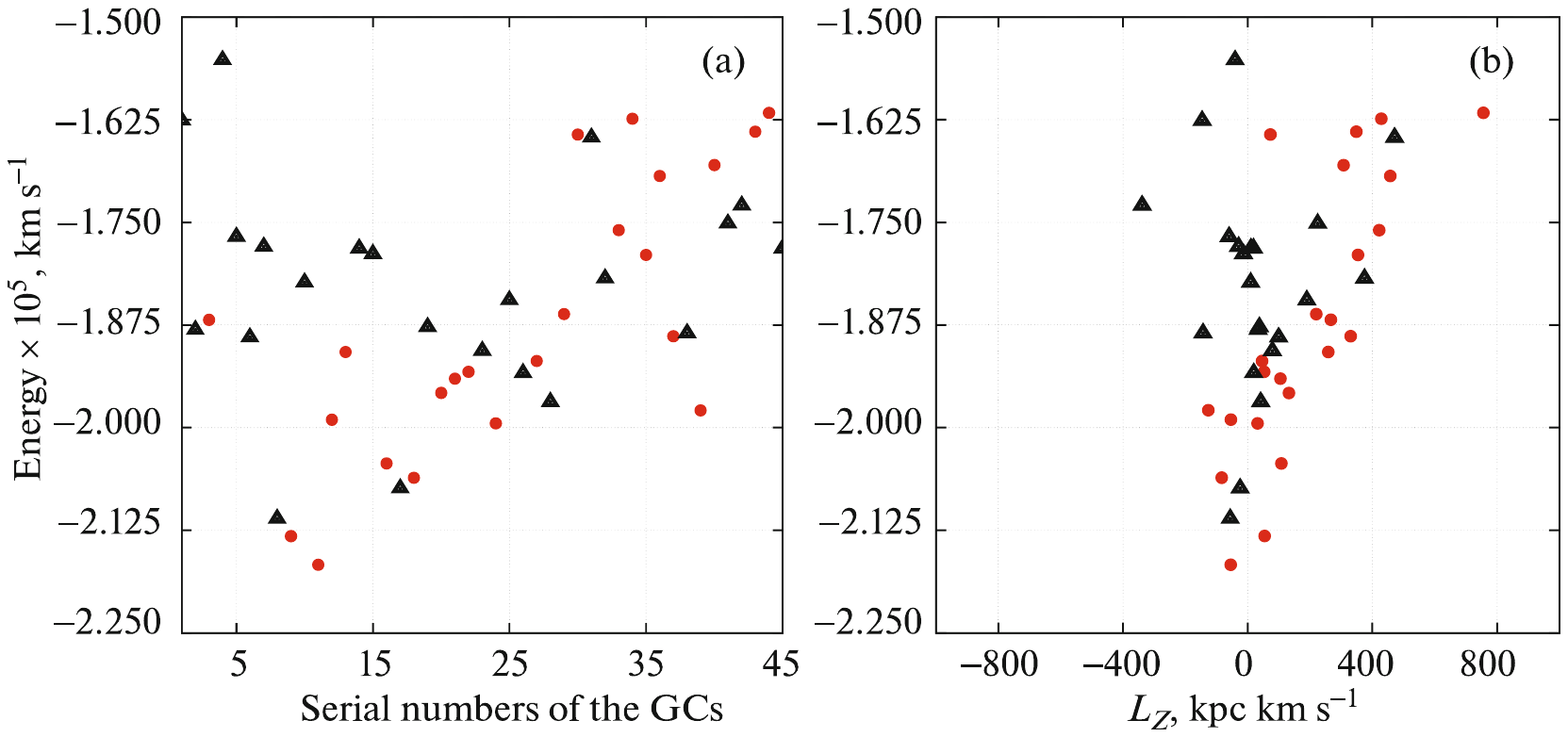}
\caption{\small Total energy at the initial moment of time depending on the serial number of the GC (a) and the diagram "moment
$L_z$ -- total energy" (b). Red circles indicate GCs with regular orbits and black triangles -- with chaotic orbits.}
\label{Lyap72}
\end{center}}
\end{figure*}

\begin{figure*}
{\begin{center}
               \includegraphics[width=0.9\textwidth,angle=0]{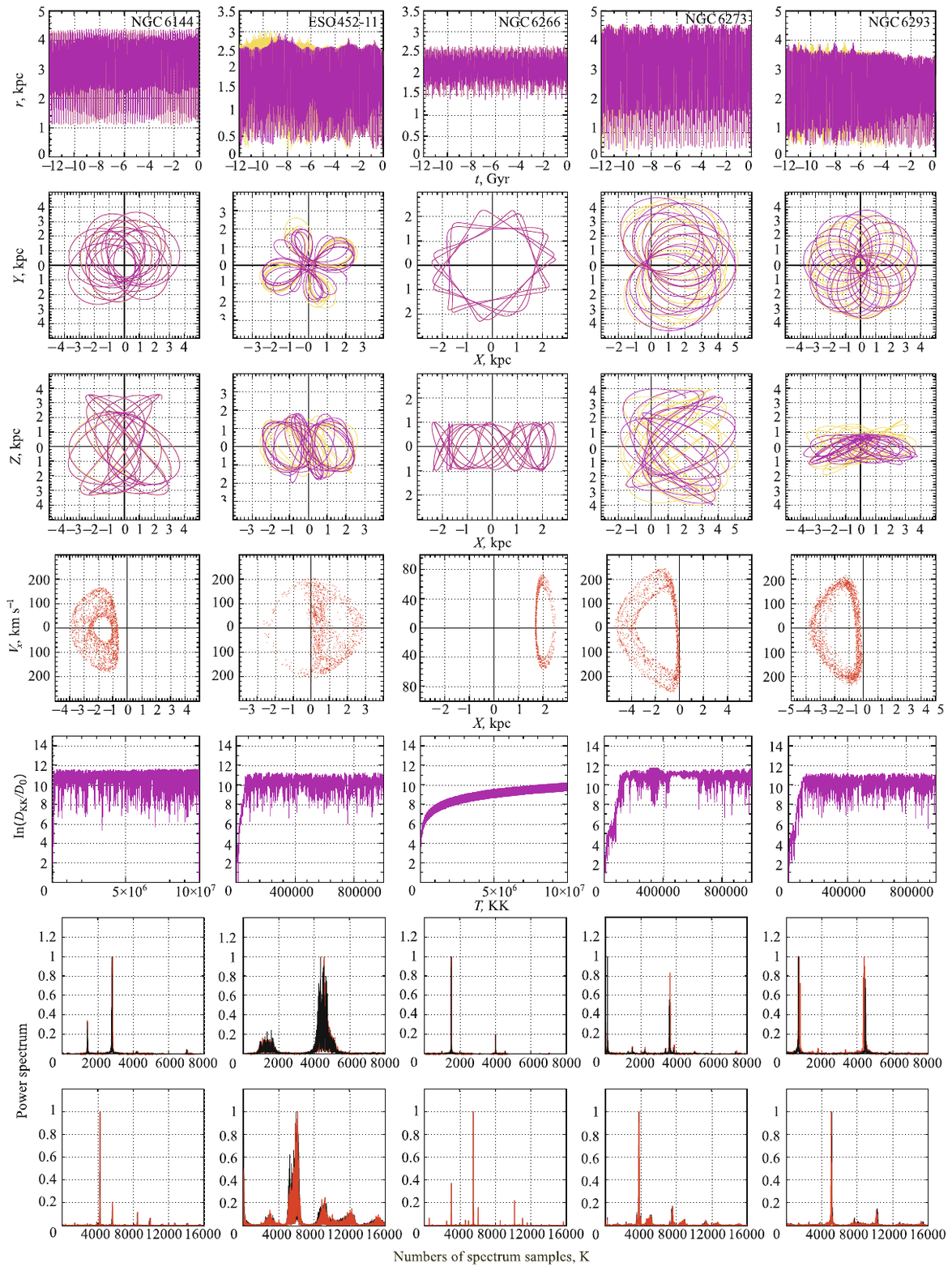}
\caption{\small Orbits of globular clusters. In the panels from top to bottom: radial values of the orbit depending on time, "X -- Y" orbit projection, "X -- Z" projection (the reference orbit is shown in yellow, the shadow orbit in purple);
Poincar\'e section "X -- $V_x$"; illustration of the probabilistic method; illustration of the frequency method (the amplitude spectrum of the first half of the time sequence is shown in red, the second half in black); normalized amplitude spectra of the radial values of the reference and shadow orbits as functions of time, shown in black and red, respectively. The names of the GCs are indicated in the panels of the top row.}
\label{Lyap13}
\end{center}}
\end{figure*}

\begin{figure*}
{\begin{center}
               \includegraphics[width=1.0\textwidth,angle=0]{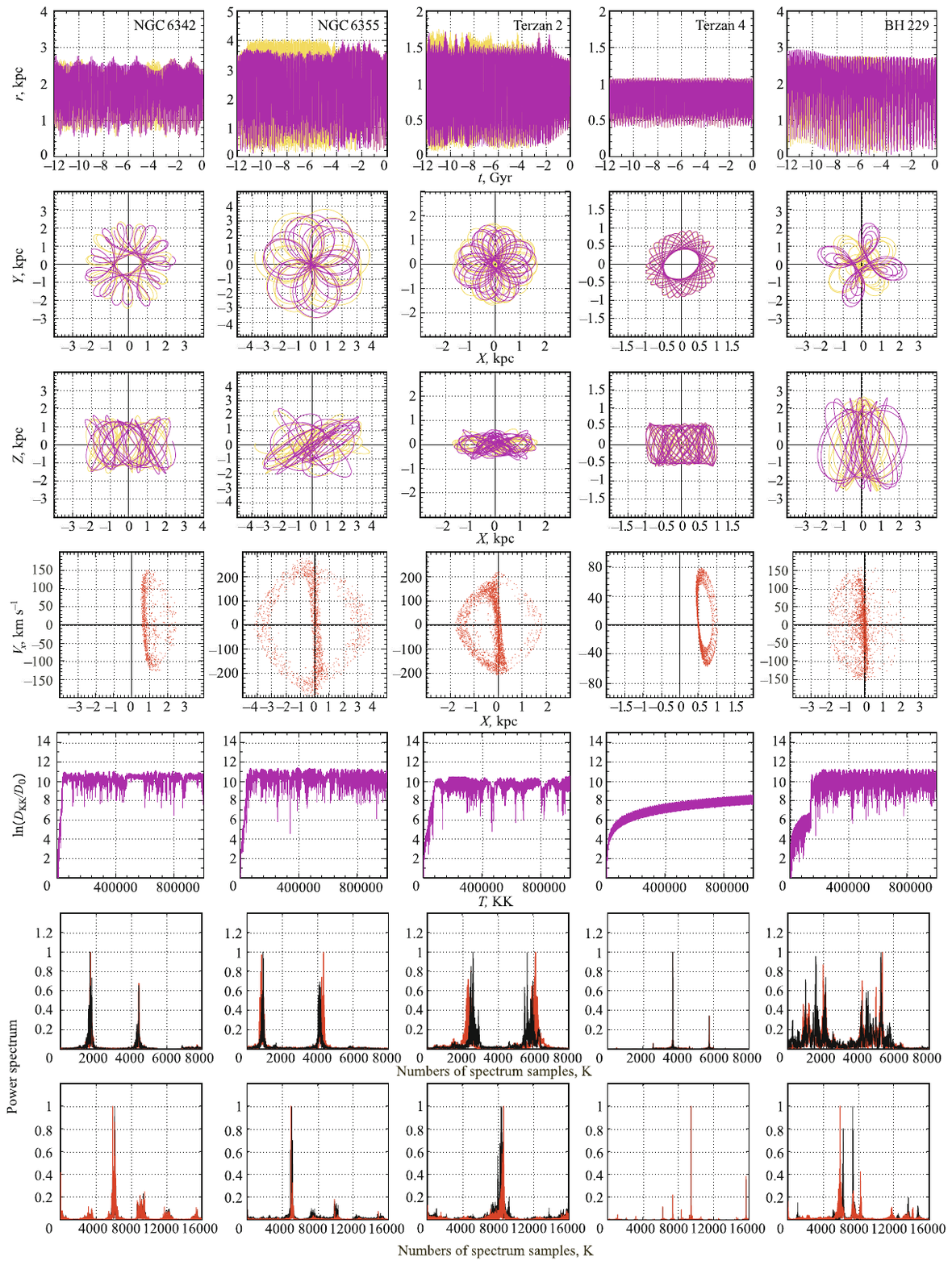}
\centerline{Figure 13: Contd.}
\label{Lyap13}
\end{center}}
\end{figure*}

\begin{figure*}
{\begin{center}
               \includegraphics[width=1.0\textwidth,angle=0]{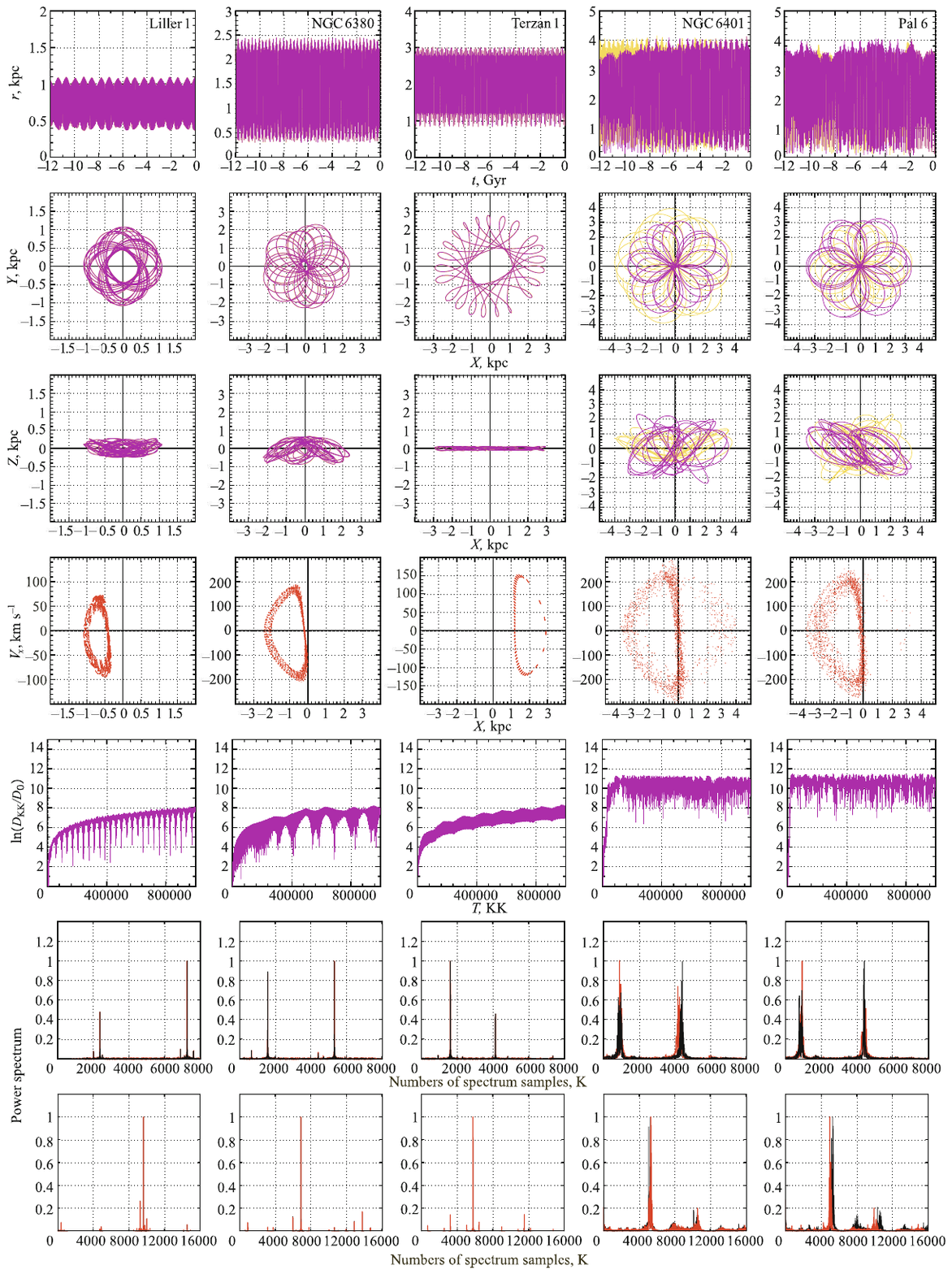}
\centerline{Figure 13: Contd.}
\label{Lyap13}
\end{center}}
\end{figure*}

\begin{figure*}
{\begin{center}
               \includegraphics[width=1.0\textwidth,angle=0]{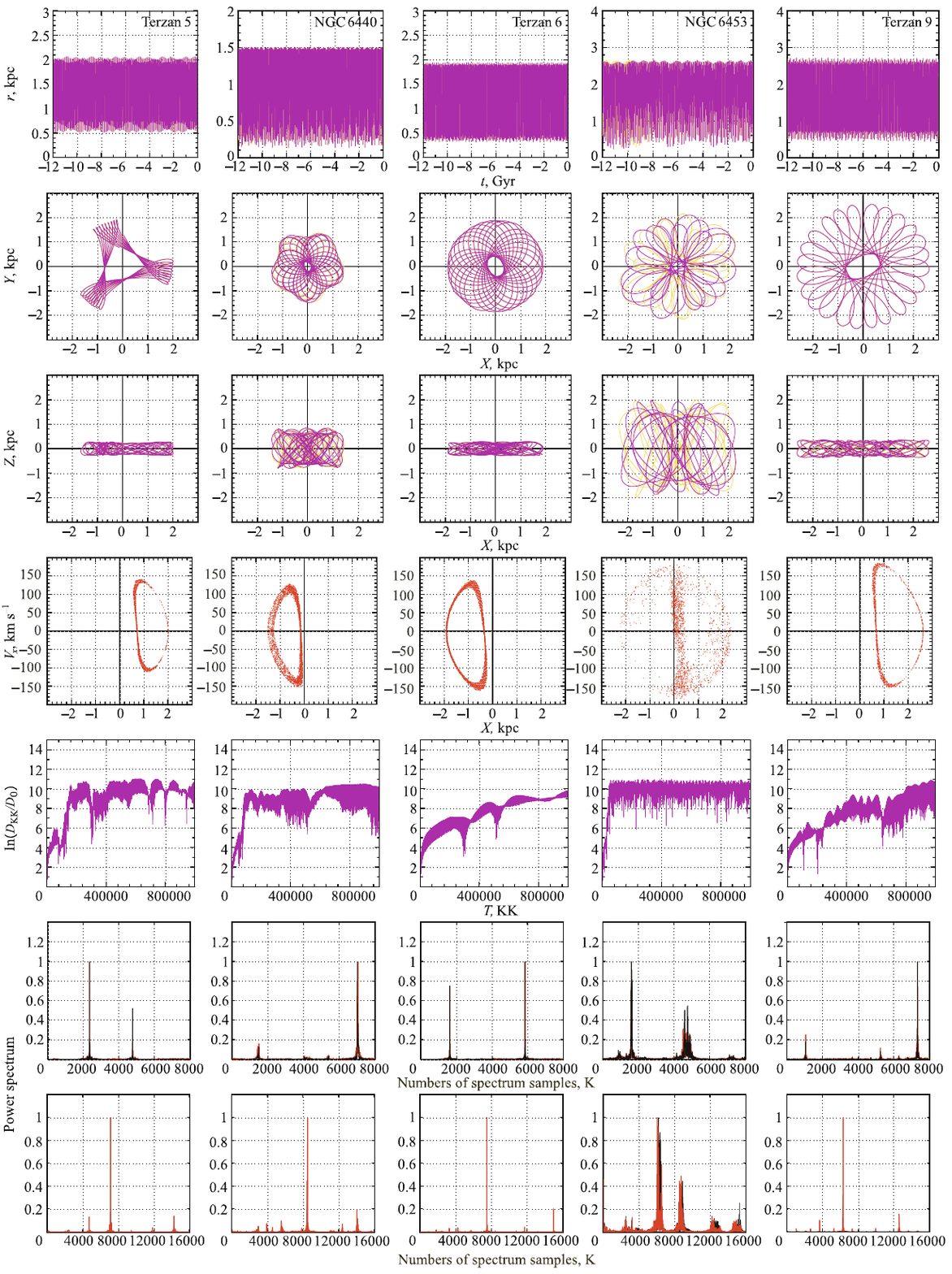}
\centerline{Figure 13: Contd.}
\label{Lyap13}
\end{center}}
\end{figure*}

\begin{figure*}
{\begin{center}
               \includegraphics[width=1.0\textwidth,angle=0]{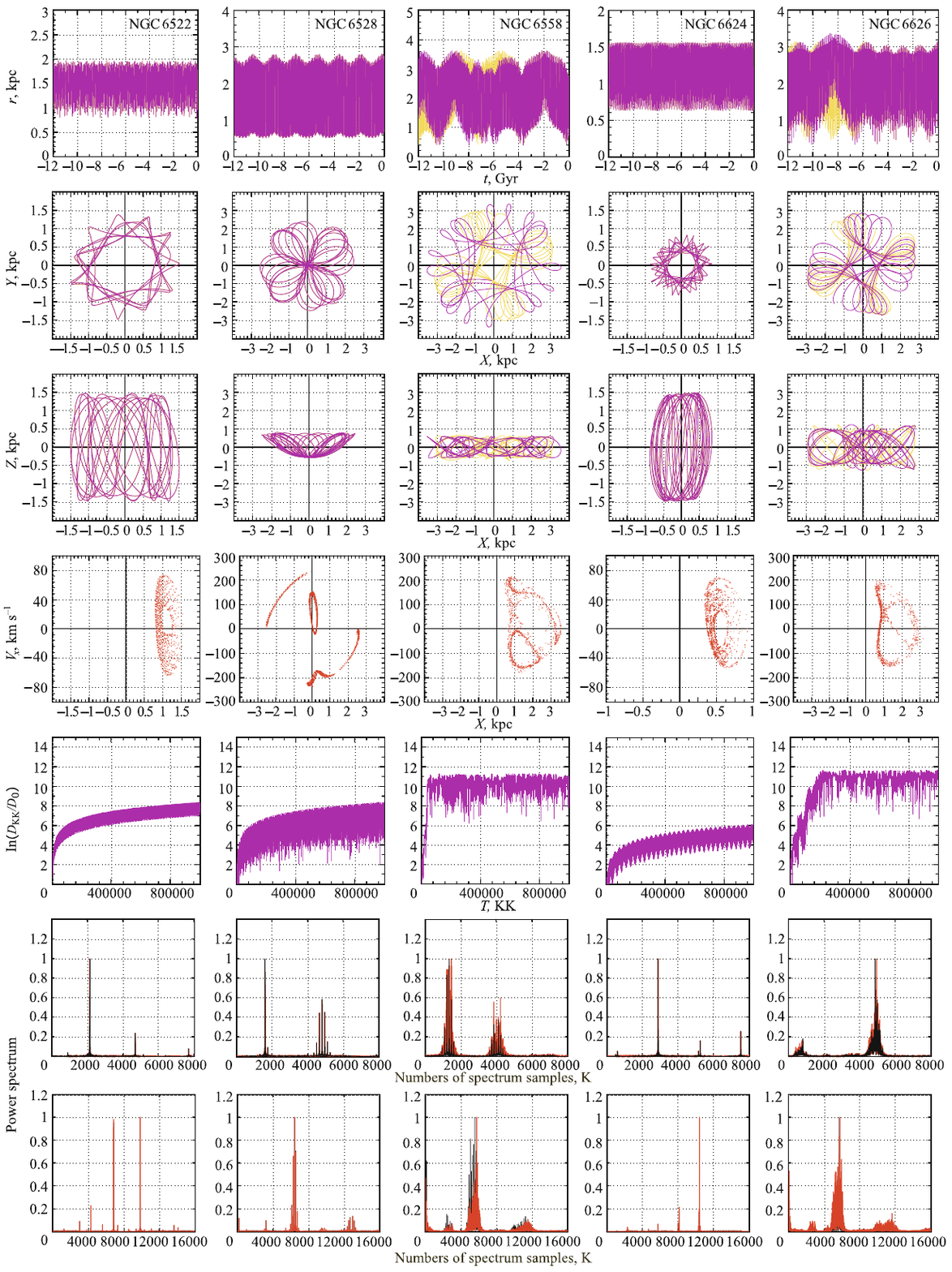}
\centerline{Figure 13: Contd.}
\label{Lyap13}
\end{center}}
\end{figure*}

\begin{figure*}
{\begin{center}
               \includegraphics[width=1.0\textwidth,angle=0]{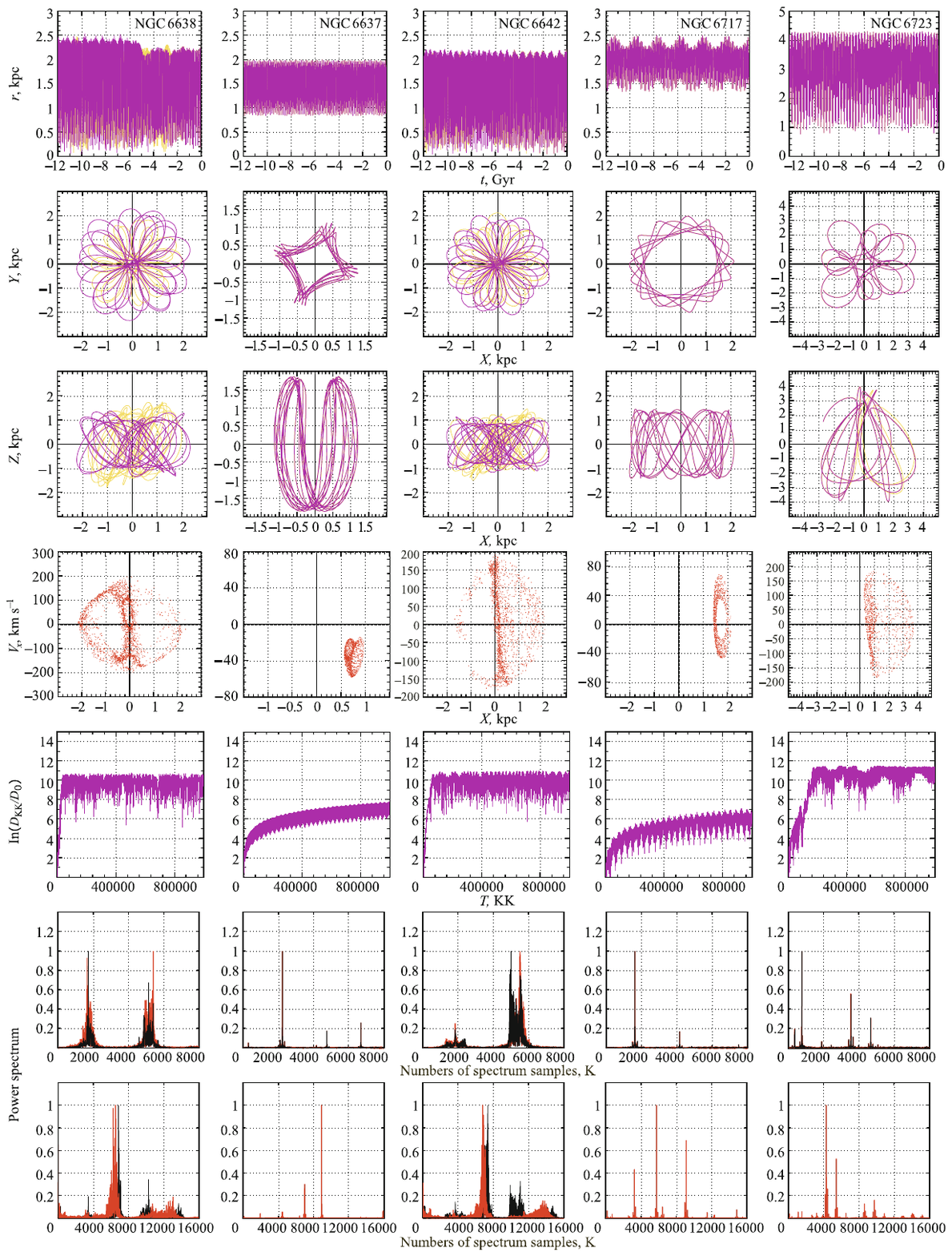}
\centerline{Figure 13: Contd.}
\label{Lyap13}
\end{center}}
\end{figure*}

\begin{figure*}
{\begin{center}
               \includegraphics[width=1.0\textwidth,angle=0]{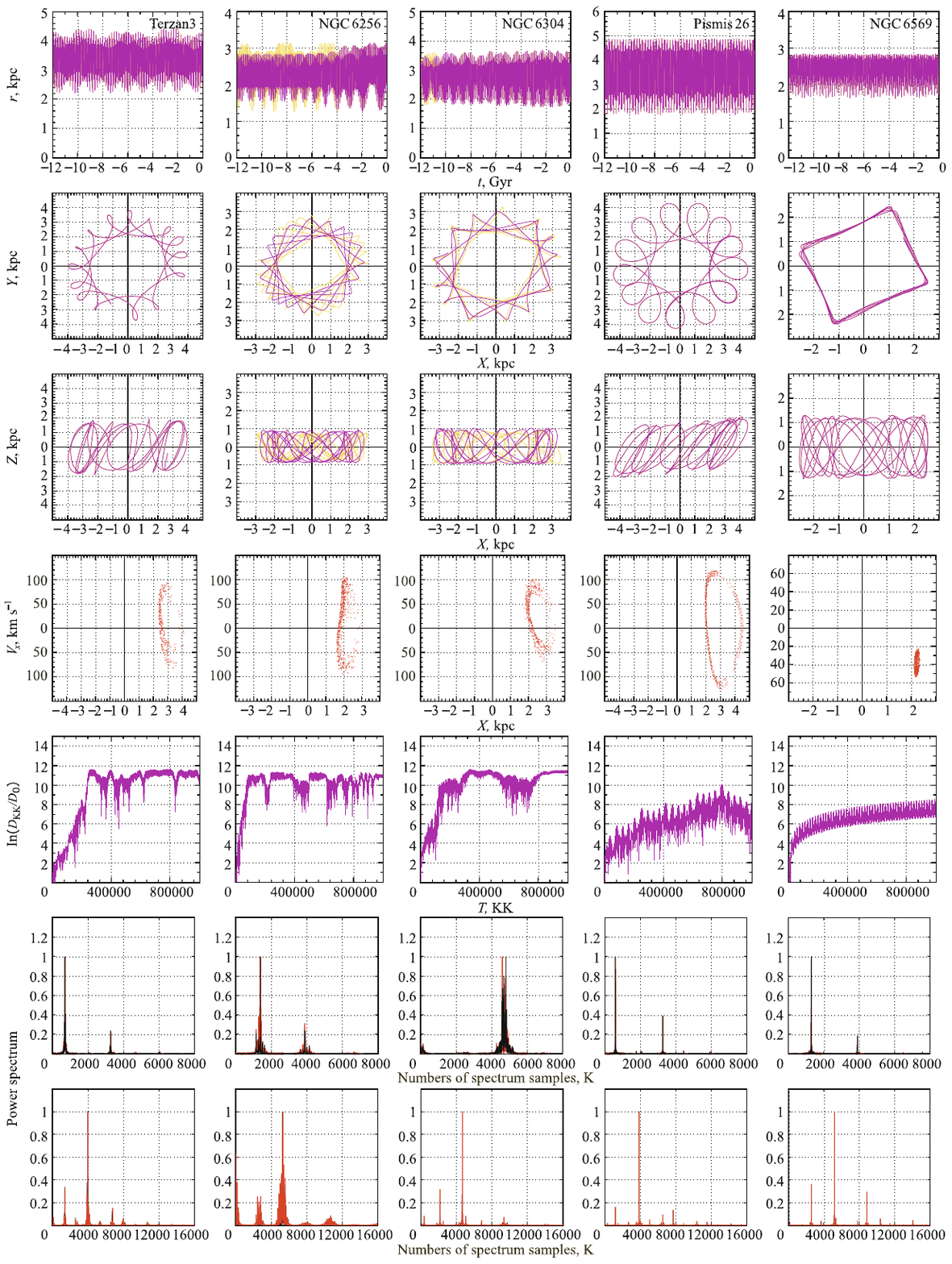}
\centerline{Figure 13: Contd.}
\label{Lyap13}
\end{center}}
\end{figure*}

\begin{figure*}
{\begin{center}
               \includegraphics[width=1.0\textwidth,angle=0]{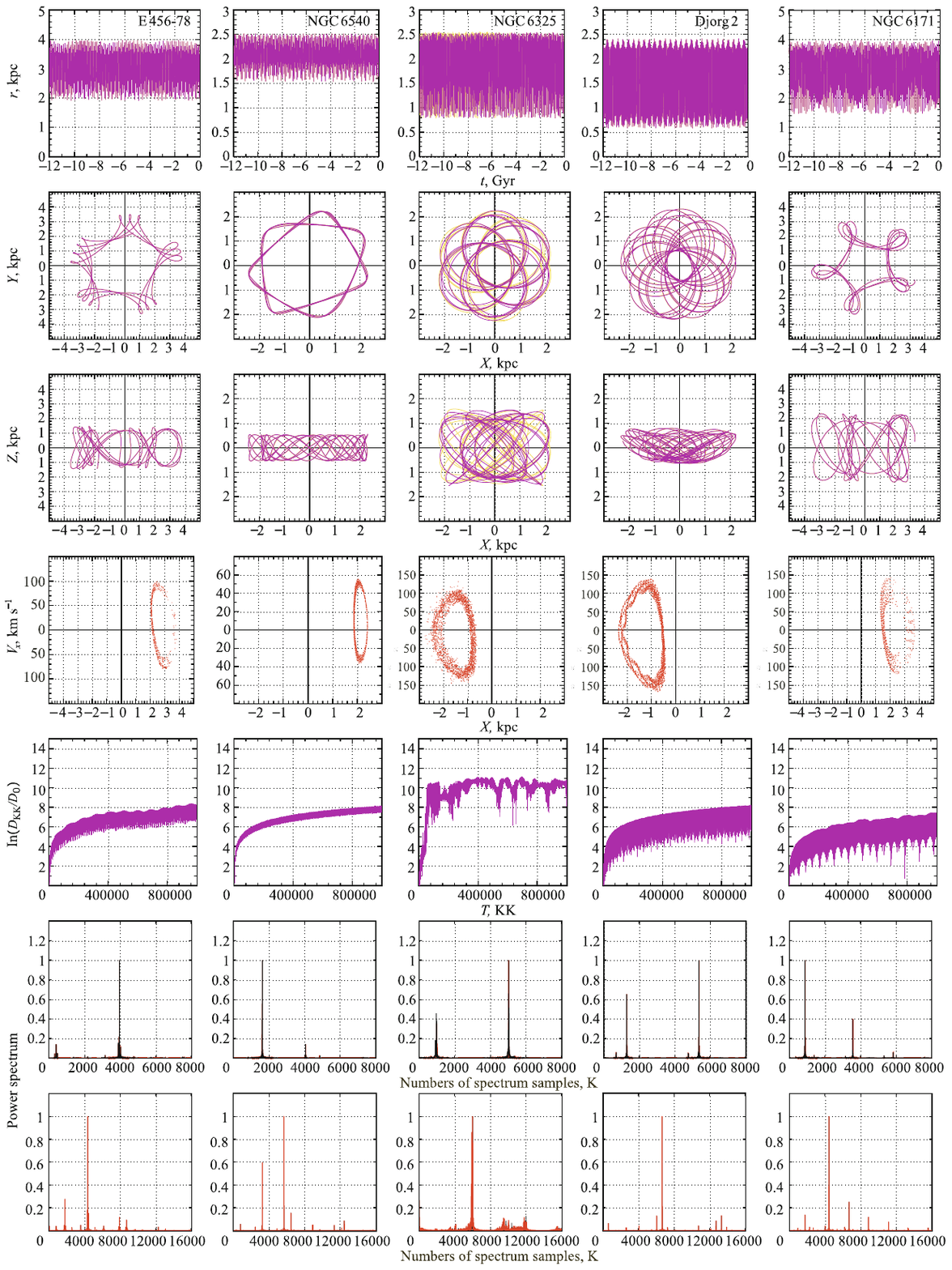}
\centerline{Figure 13: Contd.}
\label{Lyap13}
\end{center}}
\end{figure*}

\begin{figure*}
{\begin{center}
               \includegraphics[width=1.0\textwidth,angle=0]{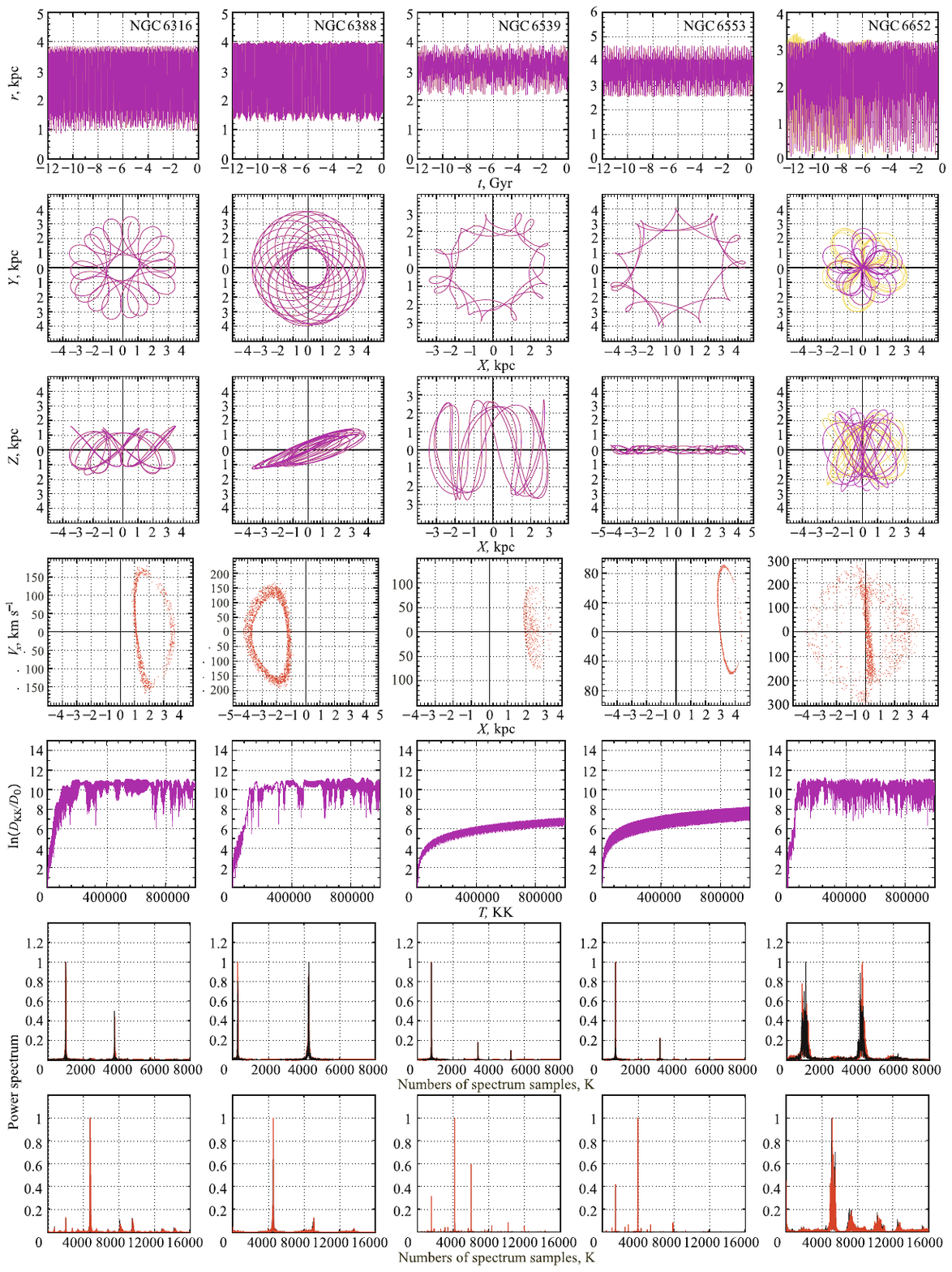}
\centerline{Figure 13: Contd.}
\label{Lyap13}
\end{center}}
\end{figure*}

\begin{figure*}
{\begin{center}
               \includegraphics[width=1.0\textwidth,angle=0]{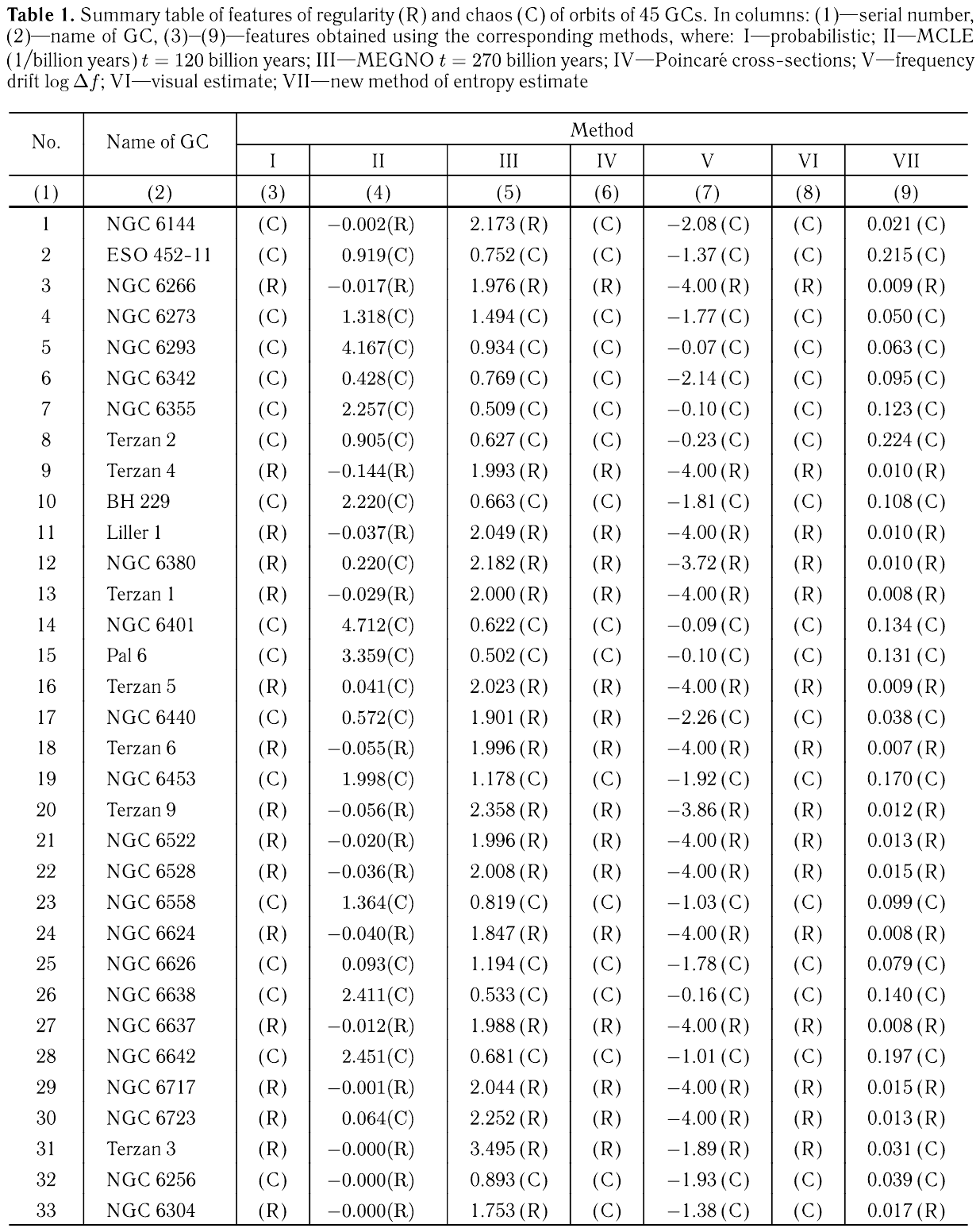}
\end{center}}
\end{figure*}

\begin{figure*}
{\begin{center}
               \includegraphics[width=0.7\textwidth,angle=-90]{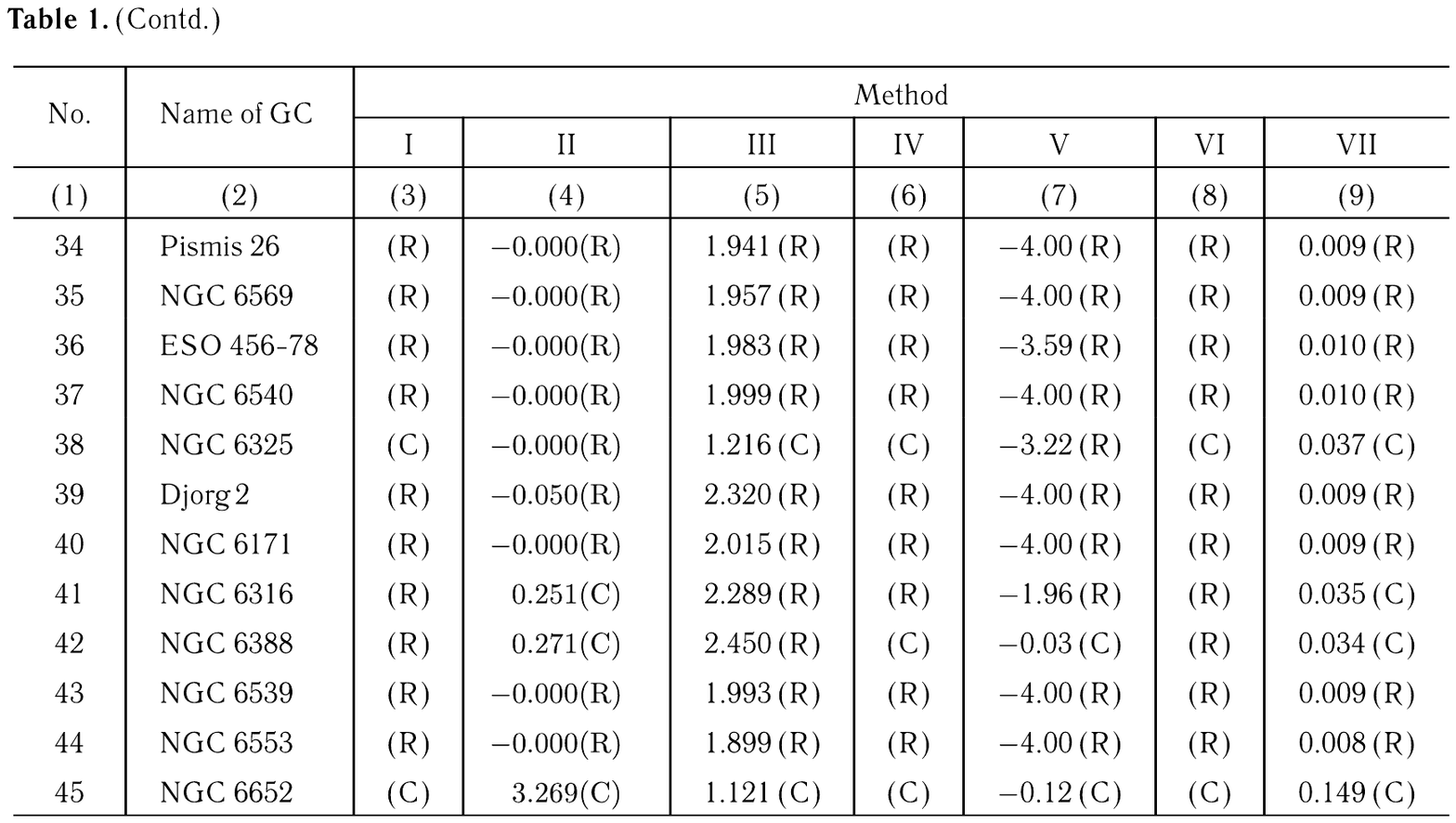}
\end{center}}
\end{figure*}

\begin{figure*}
{\begin{center}
               \includegraphics[width=0.4\textwidth,angle=-90]{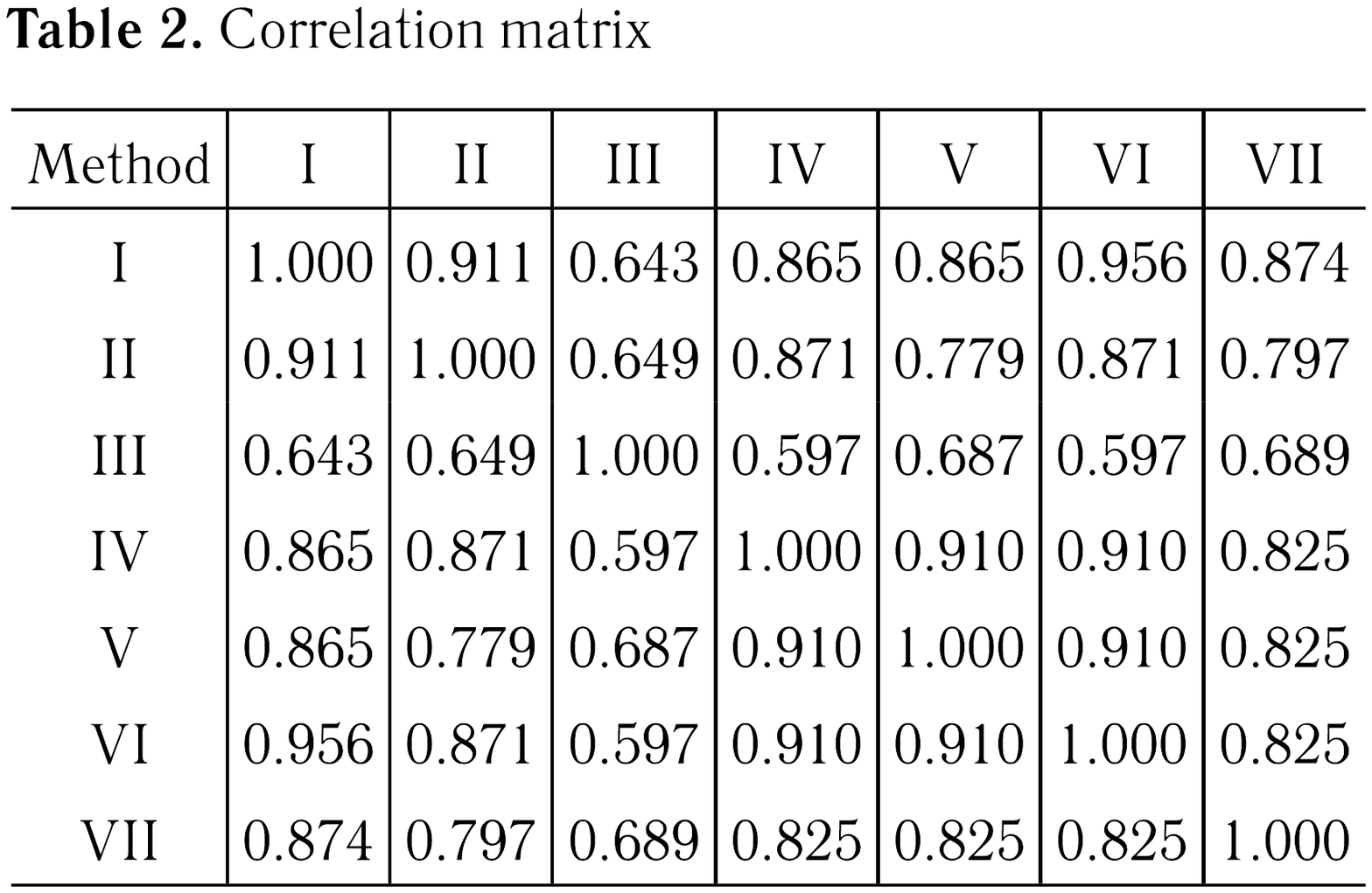}
\end{center}}
\end{figure*}

\section{CONCLUSIONS}
For the first time, the regularity/chaoticity of the
orbital dynamics of a sample of 45 globular clusters
in the central region of the Milky Way with a radius
of 3.5 kpc was studied in a non-axisymmetric
gravitational potential (Bajkova et al., 2023a) with
an elongated rotating bar. The bar model used is a
triaxial ellipsoid with a mass of $10^{10} M_\odot$, a semi-major
axis length of 5 kpc, an inclination angle to
the galactic $X$ axis of 25$^o$, and a rotation velocity of
40 km s$^{-1}$ kpc$^{-1}$. The orbits were constructed using
the most accurate astrometric data to date from the
Gaia EDR3 satellite (Vasiliev and Baumgardt, 2021),
as well as new refined average distances to globular
clusters (Vasiliev and Baumgardt, 2021).

As a result of the initial calculation of the maximum
characteristic Lyapunov exponents by a direct
method (i.e. without renormalization of the shadow
trajectory of the GC), bimodality was detected in the
histogram of the distribution of approximations of
the Lyapunov exponents. Using a method based on
the approximation of the histogram by two Gaussian
probability distributions, the GC sample was divided
into objects with regular and chaotic dynamics. An
explanation of the detected bimodality is given. The
proposed method is called probabilistic and essentially
represents a new approach.

In addition to the probabilistic method, several
other well-known methods were used to analyze GCs
for regularity/chaoticity, namely: the MCLE calculation
with shadow orbit renormalization, MEGNO,
the Poincar\'e cross-section method, the frequency
method based on the estimation of the fundamental
frequency drift parameter, and the method of visual
estimation of orbits on long time intervals comparable
to the age of the Universe. All the methods used were
shown to correlate well with each other. The final
decision on classifying GCs as objects with regular or
chaotic dynamics was made by the "voting" principle,
taking into account the results of the analysis of all
the methods used. Thus, two lists of globular clusters
were formed:

1) 24 objects with regular dynamics:
NGC6266, Terzan4, Liller1, NGC6380,
Terzan1, Terzan5, Terzan6, Terzan9,
NGC6522, NGC6528, NGC6624, NGC6637,
NGC6717, NGC6723, Terzan3, Pismis26,
NGC6569, ESO456-78, NGC6540, Djorg2,
NGC6171, NGC6316, NGC6539, NGC6553;

2) 21 objects with chaotic dynamics:
NGC6144, ESO452-11, NGC6273, NGC6293,
NGC6342, NGC6355, Terzan2, BH229, Pal6,
NGC6401, NGC6440, NGC6453, NGC6558,
NGC6626, NGC6638, NGC6642, NGC6256,
NGC6304, NGC6325, NGC6388, NGC6652.

A new simple method for determining the nature
of orbital motion, chaotic or regular, is proposed. It
is based on calculating the amplitude spectrum of
the orbit as a function of time and the entropy of the
amplitude spectrum as a measure of orbital chaos.
Based on the new method, a list of 23 GCs with
regular dynamics is determined: NGC 6266, Terzan 4, Liller 1, NGC 6380,
Terzan1, Terzan5, Terzan6, Terzan9,
NGC6522, NGC6528, NGC6624, NGC6637,
NGC6717, NGC6723, NGC6304, Pismis26,
NGC6569, ESO456-78, NGC6540, Djorg2,
NGC6171, NGC6539, NGC6553, and
a list of 22 GCs with chaotic dynamics:
NGC6144, ESO452-11, NGC6273, NGC6293,
NGC6342, NGC6355, Terzan2, BH229,
NGC6401, Pal6, NGC6440, NGC6453,
NGC6558, NGC6626, NGC6638, NGC6642,
Terzan3, NGC6256, NGC6325, NGC6316,
NGC6388, NGC6652.

Comparison of the classification results of the GCs
obtained by the proposed method and known methods
revealed a high correlation equal to 0.825.

As shown by the analysis of the parameters of the
reference and shadow orbits on time intervals comparable
with the age of the Universe, the list of GCs with
chaotic dynamics included mainly globular clusters
with elongated radial orbits with apocentric distances
in the region of 2.5 -- 3.5 kpc, with small pericentric
distances (less than 0.4 kpc) and large eccentricities
(more than 0.8) (the correlation between the last
orbital parameters was 0.85), which is explained by
the influence of the rotating bar on the dynamics of
such GCs to the greatest extent, ultimately leading to
chaotic motion.

\bigskip

\noindent{\bf ACKNOWLEDGMENTS}

\bigskip

The authors are grateful to A.V. Melnikov for discussing
the problem of orbital stability. The authors
are grateful to the anonymous reviewers for
their careful reading of the article and interesting
comments that will serve as a stimulus for further
research.


\begin{thebibliography}{25}
\providecommand{\natexlab}[1]{#1}
\bibitem
1~A. T. Bajkova, A. A. Smirnov, and V. V. Bobylev, Publications
of the Pulkovo Observatory {\bf 228}, 1 (2023a).
DOI:10.31725/0367-7966-2023-228-1
\bibitem
2~A. T. Bajkova, A. A. Smirnov, and V. V. Bobylev,
Astrophysical Bulletin {\bf 78} (4), 499 (2023c).
https://doi.org/10.1134/S199034132360028X
\bibitem
3~H. Baumgardt and E.Vasiliev, MonthlyNotices Royal
Astron. Soc. {\bf 505} (4), 5957 (2021).
https://doi.org/10.1093/mnras/stab1474
\bibitem
4~V. V. Bobylev and A. T. Bajkova, Publications of the
Pulkovo Observatory {\bf 227}, 36 (2022).
https://doi.org/10.31725/0367-7966-2022-227-3
\bibitem
5~S. Breiter, B. Melendo, P. Bartczak, and
I. Wytrzyszczak, Astron. and Astrophys. {\bf 437} (2), 753
(2005).
https://doi.org/10.1051/0004-6361:20053031
\bibitem
6~O. V. Chumak, {\it Entropies and Fractals in Data
Analysis} (Moskow-Izhevsk, Institute of Computer
Science, R \& C Dynamics, 2011) [in Russian].
https://doi.org/10.13140/2.1.4739.6800
\bibitem
7~R. E. G. Machado and T. Manos, Monthly Notices
Royal Astron. Soc. {\bf 458} (4), 3578 (2016).
https://doi.org/10.1093/mnras/stw572
\bibitem
8~A. V. Melnikov, Solar System Research {\bf 52} (5), 417
(2018). https://doi.org/10.1134/S0038094618050064
\bibitem
9~A. Morbidelli, {\it Modern Celestial Mechanics. Aspects
of Solar System Dynamics} (Institute of Computer
Science, Moscow-Izhevsk, 2014) [in Russian].
\bibitem
1~K. Murray and S. Dermott, {\it Solar System Dynamics}
(Cambridge University Press, Cambridge, 1999).
https://doi.org/10.1017/CBO9781139174817
\bibitem
1~N. Nieuwmunster,M. Schultheis, M. Sormani, et al.,
Astron. and Astrophys. {\bf 685}, id. A93 (2024).
https://doi.org/10.1051/0004-6361/202349000
\bibitem
1~J. Palous, B. Jungwiert, and J. Kopecky, Astron. and
Astrophys. {\bf 274}, 189 (1993).
\bibitem
1~J. L. Sanders, L. Smith, and N. W. Evans, Monthly
Notices Royal Astron. Soc. {\bf 488} (4), 4552 (2019).
https://doi.org/10.1093/mnras/stz1827
\bibitem
1~M. Valluri, V. P. Debattista, T. Quinn, and B. Moore,
Monthly Notices Royal Astron. Soc. {\bf 403} (1), 525
(2010). https://doi.org/10.1111/j.1365-2966.2009.16192.x
\bibitem
1~E.Vasiliev and H. Baumgardt, MonthlyNotices Royal
Astron. Soc. {\bf 505} (4), 5978 (2021).
https://doi.org/10.1093/mnras/stab1475
\end{thebibliography}
\end{document}